\documentclass[sigconf,nonacm]{acmart}
\copyrightyear{2022}
\acmYear{2022}
\acmDOI{10.1145/3526113.3545650}
\acmISBN{978-1-4503-9320-1/22/10}
\acmVolume{}
\acmNumber{}
\acmConference[UIST '22]{The 35th Annual ACM Symposium on User Interface Software and Technology}{October 29-November 2, 2022}{Bend, OR, USA}
\acmBooktitle{Proceedings of the 35th Annual ACM Symposium on User Interface Software and Technology (UIST '22)}
\AtBeginDocument{
  \providecommand\BibTeX{{
    \normalfont B\kern-0.5em{\scshape i\kern-0.25em b}\kern-0.8em\TeX}}}

\usepackage{color}
\usepackage{siunitx}
\usepackage{subcaption}
\begin{document}

\title[MetamorphX: An Ungrounded 3-DoF Moment Display that Changes its Physical Properties]{MetamorphX: An Ungrounded 3-DoF Moment Display \\ that Changes its Physical Properties \\ through Rotational Impedance Control}
\titlenote{Author-prepared version of the paper published in UIST '22, Article 72, pp. 1--14. DOI: \url{https://doi.org/10.1145/3526113.3545650}. Copyright 2022 held by the owner/author(s). This work is licensed under \href{https://creativecommons.org/licenses/by-nc-nd/4.0/}{CC BY-NC-ND 4.0}.}

\author{Takeru Hashimoto}
\email{hashimoto@cyber.t.u-tokyo.ac.jp}
\affiliation{
  \institution{The University of Tokyo}
  \city{Bunkyo}
  \state{Tokyo}
  \country{Japan}
}

\author{Shigeo Yoshida}
\email{shigeo@star.rcast.u-tokyo.ac.jp}
\affiliation{
  \institution{The University of Tokyo}
  \city{Bunkyo}
  \state{Tokyo}
  \country{Japan}
}

\author{Takuji Narumi}
\email{narumi@cyber.t.u-tokyo.ac.jp}
\affiliation{
  \institution{The University of Tokyo}
  \city{Bunkyo}
  \state{Tokyo}
  \country{Japan}
}

\renewcommand{\shortauthors}{Hashimoto et al.}

\begin{abstract}

Humans can estimate the properties of wielded objects (e.g., inertia and viscosity) using the force applied to the hand.
We focused on this mechanism and aimed to represent the properties of wielded objects by dynamically changing the force applied to the hand.
We propose MetamorphX, which uses control moment gyroscopes (CMGs) to generate ungrounded, 3-degrees of freedom moment feedback.
The high-response moments obtained from CMGs allow the inertia and viscosity of motion to be set to the desired values via impedance control.
A technical evaluation indicated that our device can generate a moment with a 60-ms delay.
The inertia and viscosity of motion were varied by \SI{0.01}{kgm^2} and \SI{0.1}{Ns}, respectively.
Additionally, we demonstrated that our device can dynamically change the inertia and viscosity of motion through virtual reality applications.

\end{abstract}




\begin{teaserfigure}
  \includegraphics[width=\textwidth]{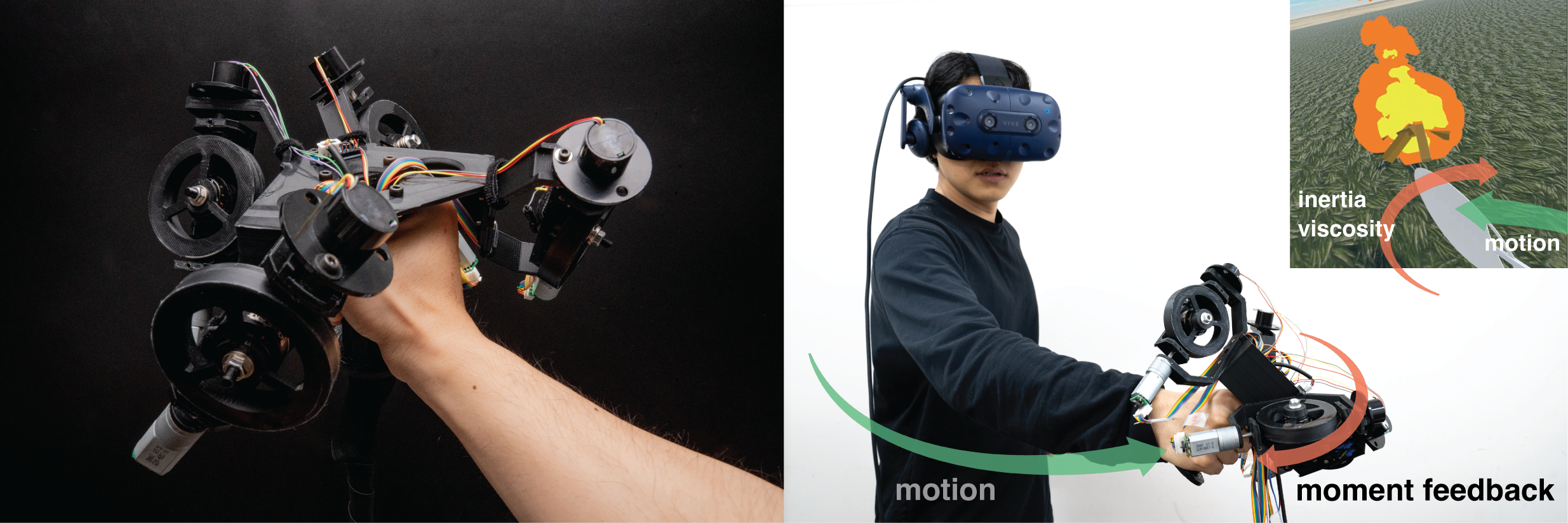}
  \caption{(Left) MetamorphX is an ungrounded 3-DoF moment display consisting of four control moment gyroscopes.
(Right) The impedance of the motion (e.g., inertia/viscosity) can be changed by the moment feedback in response to human motion. }
    \Description{(Left) MetamorphX is an ungrounded 3-DoF moment display consisting of four control moment gyroscopes.
(Right) The impedance of the motion (e.g., inertia/viscosity) can be changed by the moment feedback in response to human motion. }
  \label{fig:teaser}
\end{teaserfigure}

\maketitle

\hypersetup{
  pdftitle={MetamorphX: An Ungrounded 3-DoF Moment Display that Changes its Physical Properties through Rotational Impedance Control},
  pdfauthor={Takeru Hashimoto, Shigeo Yoshida, Takuji Narumi},
  pdfsubject={Author-prepared version; UIST 2022, Article 72, pages 1--14; DOI: 10.1145/3526113.3545650},
  pdfdoi={10.1145/3526113.3545650},
  pdfcopyright={Copyright 2022 held by the owner/author(s). Licensed under CC BY-NC-ND 4.0.},
  pdflicenseurl={https://creativecommons.org/licenses/by-nc-nd/4.0/},
  pdfpublication={Proceedings of the 35th Annual ACM Symposium on User Interface Software and Technology (UIST '22)},
  pdfpagerange={1-14},
  pdfurl={https://doi.org/10.1145/3526113.3545650}
}

\section{Introduction}

With the development of computer graphics technology and displays, audiovisual experiences in virtual reality (VR) are approaching parity with what we see and hear in the real world.
As the next step for VR, haptic feedback techniques are expected because the interaction between virtual objects and users is essential for VR.
Since haptic VR controllers play an essential role in interacting with virtual objects through a user's hand, researchers have adopted various approaches to reproduce the properties of the grasped objects.

Humans perform two types of motions with different functions for "grasping" objects \cite{napier1956prehensile}.
One is a precise grasp, in which a human grasps an object with the thumb and index finger.
The force that can be applied to the object is weak, and the human can precisely move the object.
The other is a power grasp, in which the object is held firmly by the five fingers and the whole palm.
The force that can be applied to the object is strong, and the human can dynamically move the object.

Body-grounded haptic devices have been extensively studied to represent object properties for precise grasping \cite{minamizawa2007gravity, Choi_2017, Sinclair_2019}.
Most body-grounded devices cannot reproduce the forces applied to the upper arms and shoulders.
Since most objects that humans perform precise grasping are light and small, it is not necessary to reproduce the force applied to the upper arms or shoulders to reproduce the haptic information of a precise grasp; only the reproduction of the force by the fingers is sufficient.
Hence, body-grounded devices are typically used to represent the property of an object with a precise grasp.
Various properties, e.g., weight \cite{minamizawa2007gravity,Choi_2017}, inertia \cite{minamizawa2007gravity,Choi_2017}, stiffness \cite{Sinclair_2019}, and size \cite{Choi_2017} have been represented using these body-grounded haptic devices.

However, when humans grasp objects with a power grasp, they often hold and move them using the muscles of their entire arm or the rest of the body.
Not body-grounded but ungrounded devices have been studied to reproduce the force applied to the entire arm.
Thor's hammer \cite{heo2018thor} and Aero-plane \cite{10.1145/3332165.3347926} are ungrounded force displays that generate force by controlling propellers.
By varying the force applied to the object depending on the attitude of the device, they accurately reproduce the force of gravity on the grasped object.
However, the considerable delay (approximately 300 ms) in force output makes them difficult to provide force feedback in response to high-speed human motion, as stated in \cite{heo2018thor}.

In the field of mechanical engineering, an attempt to represent the static properties of an object by generating an appropriate force output is called impedance control \cite{hogan1985impedance}.
Impedance control has the advantage that only one variable (force) is needed to represent an object's properties (e.g., inertia and viscosity).
However, impedance control requires rapid force feedback in response to human motion.
Conventional haptic devices are not capable of impedance control during power grasping owing to their delays.

Our study aimed to represent various physical properties felt from a grasped object during power grasp by impedance control using ungrounded force feedback.
Herein, we propose an ungrounded moment display called  \textit{MetamorphX}.
This device can output moments in three axial directions with low latency by utilizing the gyro effect.
By applying impedance control to the wielding motion, we attempted to present the user with the sensation of grasped objects with various physical properties.

The contributions of this study are listed below.
\begin{itemize}
    \item A novel moment display was developed that employs four CMGs enabling continuous low-latency generation of the moment in response to human motion.

    \item A technical evaluation of the proposed device showed that it could generate a moment with a 60-ms rise delay. Also, the impedance of the object could be changed, especially the inertia and viscosity of motion were varied by \SI{0.01}{kgm^2} and \SI{0.1}{Ns}, respectively.

    \item Potential applications of the device were proposed. The user study with the application indicated that the changes in inertia and viscosity could be perceived and enhance the experience of VR.

\end{itemize}

\section{Related work}

MetamorphX was inspired by various related studies in the field of haptic feedback for representing object properties.
Our idea of rotational impedance control for representing object properties is also based on studies on the human perception of grasped objects.

\subsection{Dynamic Touch}
In the field of ecological psychology, the ability of humans to perceive the properties of a grasped object through haptic perception has been investigated.
It has been shown that haptic sensation plays an essential role in the perception of the properties of a grasped object, such as the length, width \cite{Turvey1998}, and three-dimensional shape \cite{burton1990can}.
As the term ``dynamic touch'' \cite{Turvey1996-ml} suggests, all of this information is acquired by the human dynamically moving the object rather than simply holding it still.
Humans infer a plausible object shape from the forces applied to the tendons and muscles of the arm and the skin deformation of the palm by performing various actions on the object.

The approaches used to represent the properties of a grasped object can be classified into two categories: those that change the static features (passive haptic interface) and those that change the force feedback to the user (active haptic interface).

\subsection{Passive Haptic Interface}

This method represents the properties of a grasped object by changing the static physical properties of the controller.
HapTwist \cite{Zhu2019-aa} creates haptic proxies for various hand-graspable VR objects by combining low-cost artifacts.
However, this requires reassembling the proxies every time the virtual object changes.
TorqueBAR \cite{Swindells_2003}, Shifty \cite{Zenner2017}, and Transcalibur \cite{shigeyama2019transcalibur} changed the moment of inertia and center of gravity by dynamically moving a weight attached to the controller, thereby changing the perceived shape and size of the grasped object.
This eliminates the need to reassemble the equipment.
Drag:on \cite{zenner2019drag} and ShapeSense \cite{10.1145/3305367.3327991} represent the forces caused by the air resistance between the device and the environment by deforming the wings on the controller.
However, these devices are ad-hoc, requiring the use of certain mechanisms to represent certain physical properties.

\subsection{Active Haptic Interface}
Active haptic feedback reproduces the force received by a human.
Various studies have been conducted, and they can be classified into three categories according to how force is applied to the human body: environment-grounded, body-grounded, and ungrounded.

\subsubsection{Environment-grounded Force Feedback}

In an environment-grounded device, the force output structure is fixed to the environment.
For example, the multi-axis robot arm mechanism can vary the force applied to the tip of the robot arm that is in contact with the user \cite{massie1994phantom, mantis}.
The advantage is that the reaction force returns to the environment, allowing accurate force feedback to the user.
However, if the 6-DoF force and torque of a rigid body is to be reproduced, a robot arm with at least a seven DoF is required, making the robot arm larger.
Another disadvantage is that users can only interact within the reach of the robot arm.
Of course, with large robot arms, users can freely perform a wide range of interactions, but the need to install such large robot arms is unrealistic.

\subsubsection{Body-grounded Force Feedback}
A body-grounded device is one in which the force output structure is fixed somewhere on the body (e.g., the palm, arm, or wrist) and is not fixed to the environment, allowing the user to move around freely.
PIVOT \cite{Kovacs_2020} represents the weight and inertia of an object by controlling the force applied to a spherical object in the palm with an arm connected to the wrist.
CapstanCrunch \cite{Sinclair_2019} is a device grounded in the palm that reproduces the force of pinching an object with the fingertips.
It represents the size and rigidity of the grasped object.
There are also exoskeleton-type devices that cover the entire hand \cite{Gu_2016}.
Although calibration and hand-specific design are required, the fidelity of grasping an object is high because the force applied to each finger can be controlled individually.

However, a disadvantage of these body-grounded devices is that they cannot reproduce the force applied to the inner part from the grounded part, such as the upper arm or shoulder.
For example, if the device is fixed to the wrist, as in the case of PIVOT, it can reproduce the force applied to the palm; however, it cannot reproduce the force in the upper arm or shoulder.
This is not a problem when small forces are reproduced, or when users perform small motions that often occur with a precise grasp.
However, if a larger force is needed to move the grasped object, the human body attempts to use more of its muscles to generate motion, making it challenging to reproduce all the forces on the body.

\subsubsection{Ungrounded Force Feedback}

Numerous ungrounded force displays that are closely related to MetamorphX have been proposed.
Thor's hammer\cite{heo2018thor} uses the reaction force generated by pushing air.
This force represents the external force on the grasped object and the gravitational force on the grasped object.
Aero-plane \cite{10.1145/3332165.3347926} is another force display that uses the force of pushing air to represent the sensation of a weight moving in various directions on a grasped object.
However, while propeller-based devices produce large forces, there is a considerable delay before the force is output (around 300ms).
This delay is disadvantageous when the device must continuously adjust the force to human motion.

In contrast, JetController\cite{Wang2021-sd} and AirRacket \cite{Tsai2022-sz} can output impulsive force with low latency using air propulsion jets.
These devices reproduce the instantaneous impact force that occurs when a virtual grasped tool strikes a virtual object.
While the device using compressed air can present force at a high frequency, it is impossible to dynamically change air pressure with the pressure regulator they use, so it is impossible to change the force magnitude dynamically in response to human motion.

In addition, some studies tried to present forces to the user by instantly changing physical properties rather than directly generating forces.
One study used asymmetric vibrations to induce force sensation in humans\cite{4600289}.
Also, Unident \cite{Shimizu2021-df} generates an impulsive force by changing the moment of inertia of the device momentarily during swinging.
Since these methods utilize the phenomenon of human illusion, the magnitude of the presented force is not large.

\subsection{Gyro-based Ungrounded Force Feedback}
In addition to the ungrounded force feedback discussed in the previous section, researchers have studied devices utilizing changes in angular momentum.
We will discuss these devices, the area most closely related to this research, and clarify the novelty of our study.

Angular-momentum-based devices are mainly divided into reaction wheels (RWs) that change the speed of a rotating flywheel and control moment gyroscopes (CMGs) that change the attitude of the rotating flywheel using gimbals.
Both are often used to reproduce impulsive moments \cite{sakai2003effective, Winfree2009-rv, Walker2016, Walker_2018} due to their responsiveness.
CMGs can output more torque than RWs with the same mass \cite{kurokawa1998geometric}.
Researchers have proposed CMGs systems with from one \cite{Winfree2009-rv}, two \cite{Winfree2009-rv,Winfree2010-qf}, three \cite{Walker_2018} and four \cite{Walker2016} degrees of freedom, but they are all used to generate impact moments for short periods and are not used for continuous moment presentation.
In spacecraft attitude control, the original purpose of CMGs, CMGs systems with four degrees of freedom to output a given moment has been studied.
The reason for having one extra degree of freedom is to improve the singularity avoidance performance of the CMGs system.

In this research, we also aim to construct a CMGs system with four degrees of freedom and to keep feeding back the appropriate torque in response to human motion.

\subsection{Position of This Research}
Various approaches have been adopted to represent the properties of grasped objects, as mentioned previously.
The following three elements must be satisfied to represent the properties of the grasped object through force feedback.
\begin{itemize}
    \item Force feedback with minimal delay
    \item Portability through ungrounded system
    \item 3-DoF force feedback for human motion
\end{itemize}
In this study, the CMGs method—a force output method satisfying these requirements—was used.
Using this system, we verified that inertia and viscosity, typical properties that affect the swinging of an object, are changed by force feedback.

\section{MetamorphX}
\subsection{Impedance Control in Rotational Motion}
This section describes the core idea of the our study: impedance control.
The concept of impedance control is to represent object's static properties as force information.
This concept was proposed so that multi-axis robot arms can output an appropriate force when they contact with humans \cite{hogan1985impedance}.
For example, when a human applies a force to the robot arm, the robot arm can be moved with a small force if the robot attempts to move in the direction of the force.
And a human feels that the robot arm is {\it light}.
With the use of force feedback proportional to the elements of motion, such as acceleration, velocity, and displacement, a human feels as if the static properties of the robot arm have changed, rather than the robot arm is moving by itself.
In this study, the possibility of using this impedance control to describe the characteristics of a grasped object was examined.
Then, for what kind of motion can impedance control be used to represent the characteristics of the grasped object?

It was noted that the moment of inertia was dominant in the perception of grasped-object properties through dynamic touch \cite{Turvey1998}.
The moment of inertia is involved in the rotational motion.
Therefore, in this study, the impedance was controlled during rotational motion.

The following equation describes the rotational motion of a rigid body and its impedance-control-based feedback.

\begin{equation}
    I\dot\omega + D\omega + K\int\omega dt = \tau_h + \tau_{gen}
\end{equation}
\begin{equation}
    \tau_{gen} = -\Delta I\dot\omega  -\Delta D\omega  - \Delta K \int \omega dt
    \label{eq:imp_rotate}
\end{equation}
Where $I$ represents the moment of inertia, $D$ represents the viscosity coefficient, and $K$ represents the spring constant, $\omega$ represents the angular velocity of the motion.
$\tau_h$ represents the moment applied to the object by the human, and $\tau_{gen}$ represents the generated moment by the device.
By generating moments in accordance with Equation (\ref{eq:imp_rotate}), the following equation is obtained:
\begin{equation}
    (I+\Delta I)\dot\omega + (D+\Delta D)\omega + (K+\Delta K)\int\omega dt = \tau_h
    \label{eq:rot}
\end{equation}
This equation implies that the inertia, viscosity coefficient and spring constant of the motion have changed to $I+\Delta I, D+\Delta D$, and$K+\Delta K$, respectively.

What physical properties of objects do these parameters correspond to?
As mentioned previously, the moment of inertia is related to the shape, length perception, etc.
The viscosity coefficient is not an object-specific property but instead describes the behavior of an object in its environment, particularly in a fluid.
It depends on the shape of the object and density of the fluid in which the object is placed.
Therefore, the shape of the object was also perceived through this drag force.
The spring constant describes the motion that occurs when there is contact with another object in an environment.
It is a variable that represents the resistance force applied for pushing after contact.
This resistance force is perceived as the softness of the grasped object or the object in contact.

This study focused on the perception of grasped objects in a contact-free situation.
Therefore, we focus on the moment of inertia and viscosity coefficient of motion.
Of course, interaction with the external environment involving contact is also an important topic, but this study focused on the behavior of the grasped object itself.

\subsection{Hardware Implementation}

\begin{figure}
    \centering
    \includegraphics[width=\linewidth]{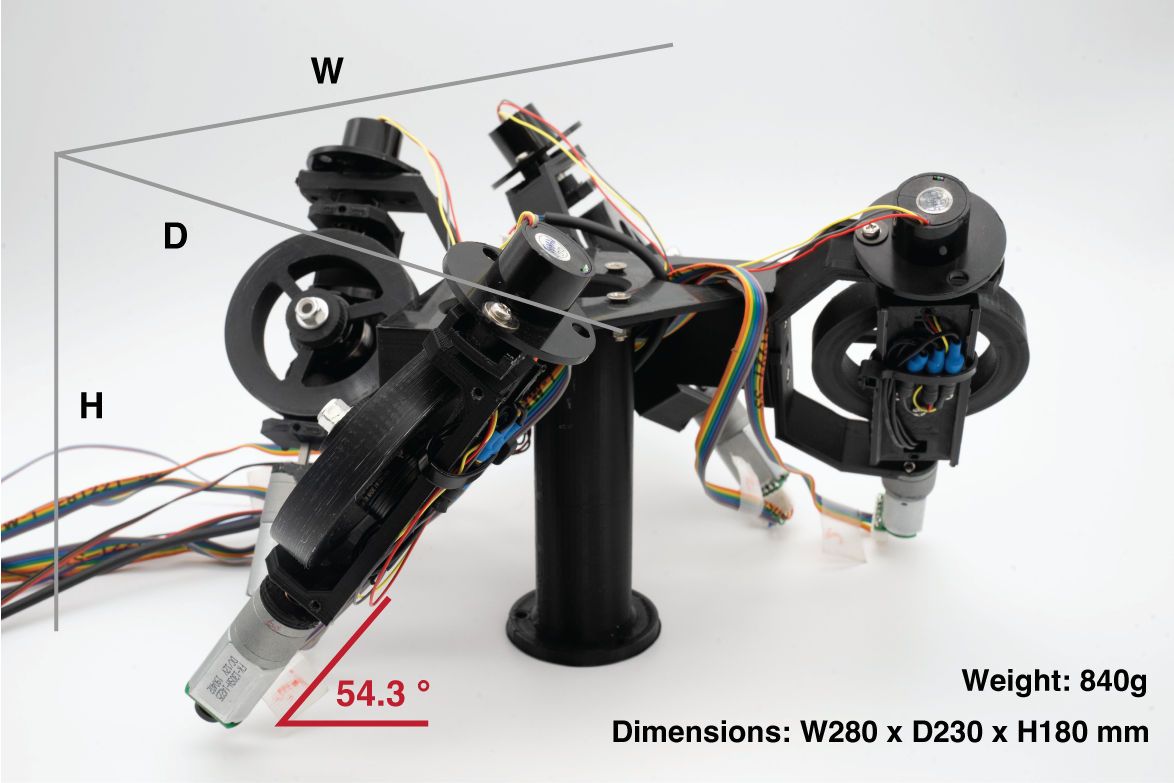}
    \caption{The overview of the MetamorphX.}
    \Description{The overview of the MetamorphX.}
    \label{fig:whole}
\end{figure}

Figure \ref{fig:whole} shows an overview of the MetamorphX, which consists of four CMGs.
For the CMGs arrangement, the four pyramid-type arrangement was selected because it can output moments evenly in the three axial directions.
The four pyramid-type arrangement was achieved by rotating and arranging each gimbal every 90 degrees with each gimbal tilted by 54.3 degrees \cite{kurokawa1998geometric}, as shown in Fig. \ref{fig:whole}.
To reduce the moment of gravity around the grasping point, the center of gravity was designed to be as close to the grasping point as possible.
Furthermore, the CMGs were placed at a sufficient distance from the hand to prevent contact with the rotating flywheel when the human grasped the device.
The device specifications are also presented in Fig. \ref{fig:whole}.

\begin{figure}
    \centering
    \includegraphics[width=0.8\linewidth]{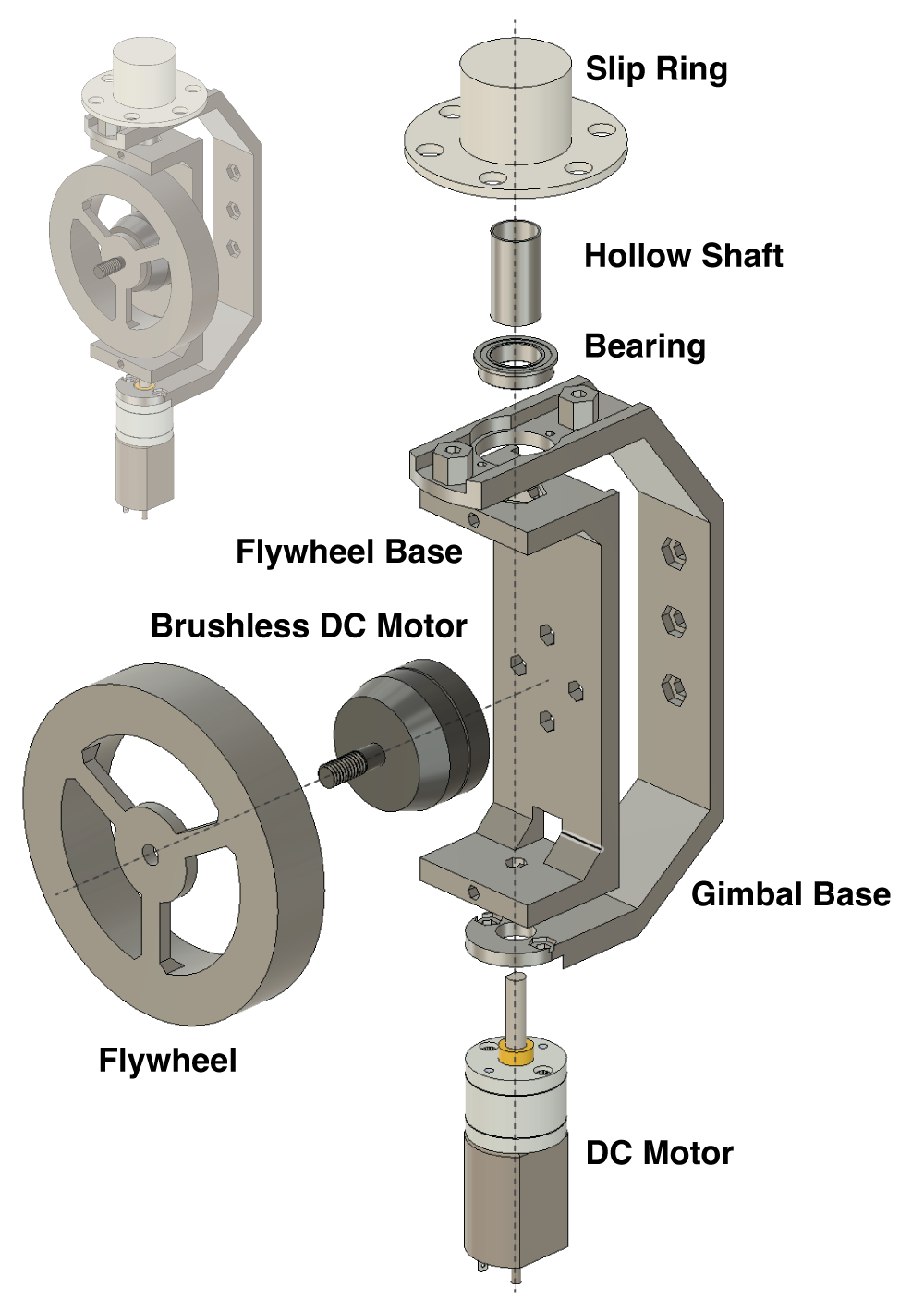}
    \caption{Exploded view of a single CMG. The brushless DC motor drives the flywheel, and the DC motor drives the flywheel base.}
    \Description{Exploded view of a single CMG. The brushless DC motor drives the flywheel, and the DC motor drives the flywheel base.}
    \label{fig:cmg}
\end{figure}

An exploded view of a single CMG is shown in Fig. \ref{fig:cmg}.
Except for mechanical parts such as bearings and slip rings, all parts were fabricated with PLA using a 3D printer (UltimakerS5).

\subsubsection{Flywheel}
The flywheel was fabricated with PLA and centered on a turning lathe to minimize the  eccentricity.
The outer diameter, the inner diameter, the mass, and the inertia were \SI{70}{mm}, 50 mm, 30 g, \SI{2.4e-5}{kgm^2}, respectively.
The parameters of the flywheel were determined so that the total mass of the device would not be heavy (less than 1 kg), and the output moment would be enough to be perceived sufficiently.
A RacerStar BR2205 sensorless brushless DC motor was used to rotate the flywheel, and a HobbyWing sensorless brushless DC speed driver was used for speed control with a PWM signal from the microncontroller (Teensy 4.0).
In this CMGs system, the rotation speed of the flywheel was constant (800 rad/s).

\subsubsection{Gimbal}
Pololu 20D metal gearmotor 63:1 was selected as the gimbal motor capable of rotating the flywheel, as described above.
The magnetic encoder attached to the motor transmitted its position and speed to a microcontroller, and the motor was driven by the motor driver (MDD10A, Cytron) with a PWM signal from the microncontroller.
The motor speed was controlled using PID controller, and the control frequency is 500 Hz.
A slip ring was installed on the opposite side of the motor to achieve infinite rotation.
This eliminated the cable entanglement caused by rotation.

\subsubsection{Inertial Measurement Unit (IMU) }
An MPU9250 was used to measure the attitude, angular velocity, and angular acceleration of the device.
The MPU9250 sends information to the microcontroller at 500 Hz.
The attitude and angular velocity were obtained directly from the sensor; however, the angular acceleration had to be calculated.
The value obtained by the difference between the angular velocity of the current frame $\omega[t]$ and the previous frame $\omega[t-\Delta t]$ divided by the interval $\Delta t$ was used as the angular acceleration value.
\begin{equation}
    \dot{\omega}[t] = \frac{\omega[t] - \omega[t-\Delta t]}{\Delta t}
\end{equation}

\subsubsection{System Diagram}

\begin{figure}
    \centering
    \includegraphics[width=\linewidth]{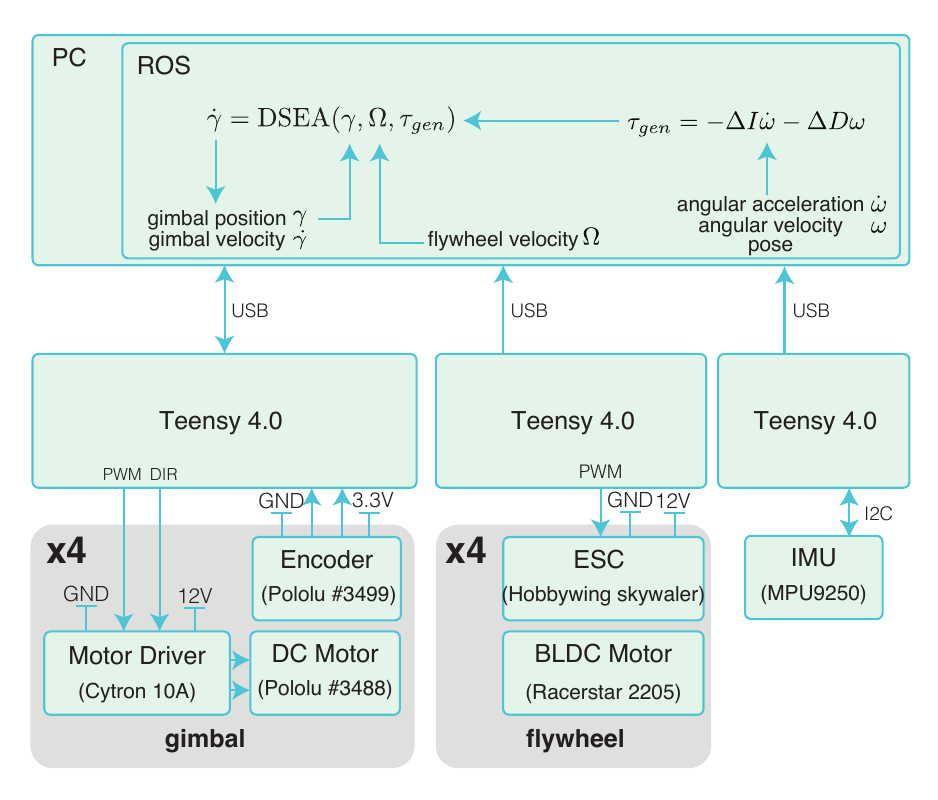}
    \caption{Block diagram of the system showing the mechatronic components and relevant connections. For simplicity, one of the four identical CMGs is shown.}
    \Description{Block diagram of the system showing the mechatronic components and relevant connections. For simplicity, one of the four identical CMGs is shown.}
    \label{fig:soft}
\end{figure}

Figure \ref{fig:soft} shows a block diagram of the system architecture.
The flywheel, gimbals, and IMU were connected to a personal computer (PC) via Teensy 4.0 microcontrollers.
The parameters obtained from each microcontroller were shared by the ROS server running on the PC.
Although it is possible to handle the gimbals, flywheels, and IMU with a single Teensy 4.0, we have divided the system into each components in this study to make it easy to handle.

Many algorithms have been proposed for calculating the gimbal velocity from the current gimbal state when a reference moment is applied.
We selected the DSEA method \cite{valk2018directional} in this study because of its stability.
It calculates the velocity of the gimbal from the moment to be output, the position of the gimbal, and the speed of the flywheel.
Additional details of DSEA were presented in the original paper \cite{valk2018directional}.

\section{Technical Evaluation}
We examined whether MetamorphX can change the impedance (inertia and viscosity) of motion through impedance control.
The following elements are needed to change the impedance of motion.
\begin{enumerate}
    \item The device can generate the specified moment with a short delay.
    \item The device can generate high-frequency moment.
    \item The device can generate moments proportional to the angular velocity and angular acceleration obtained from the IMU.
\end{enumerate}

We verify (1) and (2) in Section \ref{sec:eval_fix} and (3) in Section \ref{sec:evalimp}.

\subsection{Evaluation in the Fixed State: Delay, Frequency and Maximum Moment Impulse}\label{sec:eval_fix}
\begin{figure}
    \centering
    \includegraphics[width=0.9\linewidth]{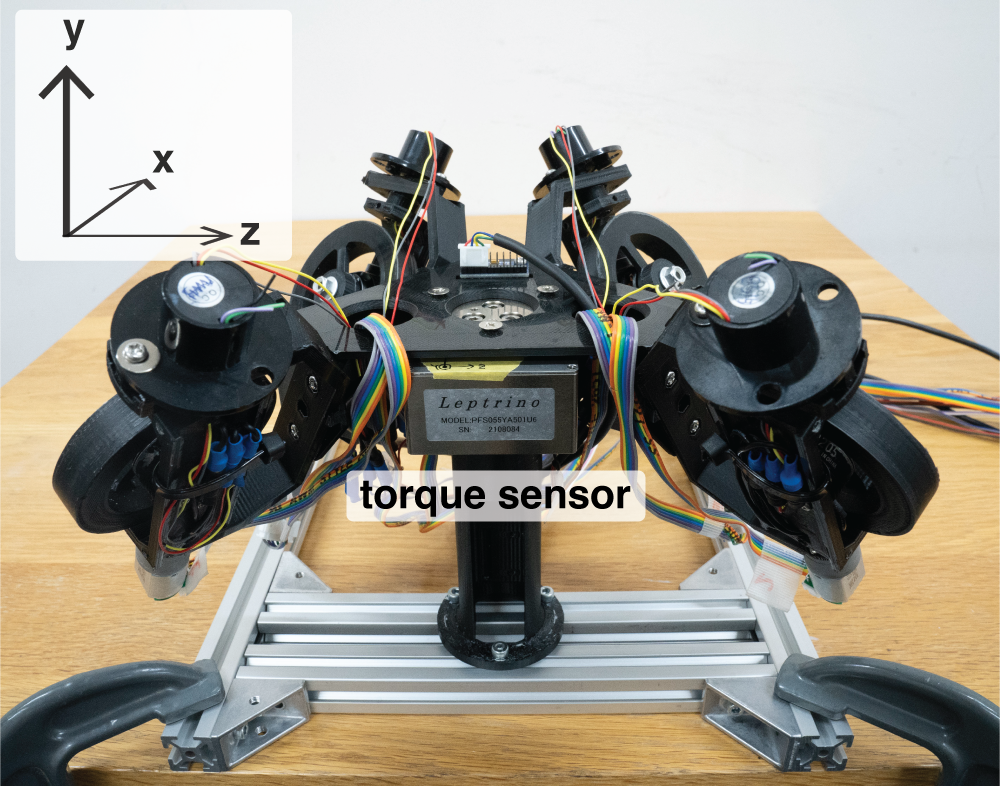}
    \caption{Setup for evaluation in fixed state. The torque sensor is mounted between moment output section and handle. The coordinate of the system is noted.}
    \Description{Setup for evaluation in fixed state. The torque sensor is mounted between moment output section and handle. The coordinate of the system is noted.}
    \label{fig:fixed}
\end{figure}

By providing a constant moment as a target value and measuring the output moment, the delay and the accuracy of the output moment were measured.
As shown in the Fig.\ref{fig:fixed}, a torque sensor (PFS055YA501U6, Leptrino) is attached between the handle and the moment output section to measure the moment.
The torque sensor had a built-in low-pass filter, and the sensor value was filtered at a cutoff frequency of 100 Hz.
The handle was fixed to an aluminium frame as Fig.\ref{fig:fixed}.

\begin{figure}[h]
    \centering
    \includegraphics[width=0.9\linewidth]{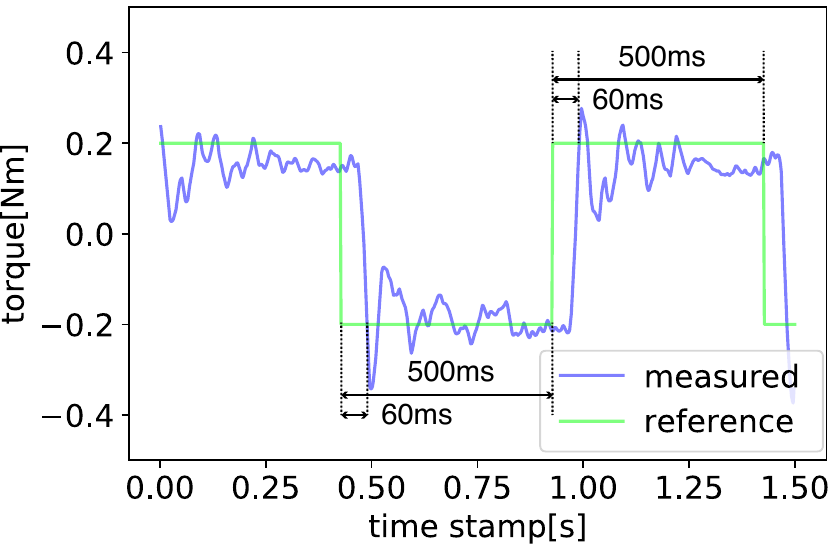}
    \caption{Response to the square wave}
    \Description{Response to the square wave}
    \label{fig:square}
\end{figure}

A square wave with an amplitude of 0.2 \si{Nm} in the x-axis was used as the target moment.
The responses are presented in Fig.\ref{fig:square}.
As shown, it took 60 milliseconds for the moment to rise.

Next, a sinusoidal moment was used as a target value to determine the frequency of moment that MetamorphX could output.
The authors of \cite{plasqui2013daily} asserted that the frequency range of human daily activities was 0.3 - 3.5 Hz.
Thus, a signal of 0.4 Nm at 0.5, 1, 2,or 4 Hz in the x-axis was used as an input.

\begin{figure}[htbp]
    \begin{tabular}{cc}
      \begin{minipage}[t]{0.45\hsize}
        \centering
        \includegraphics[width=\linewidth]{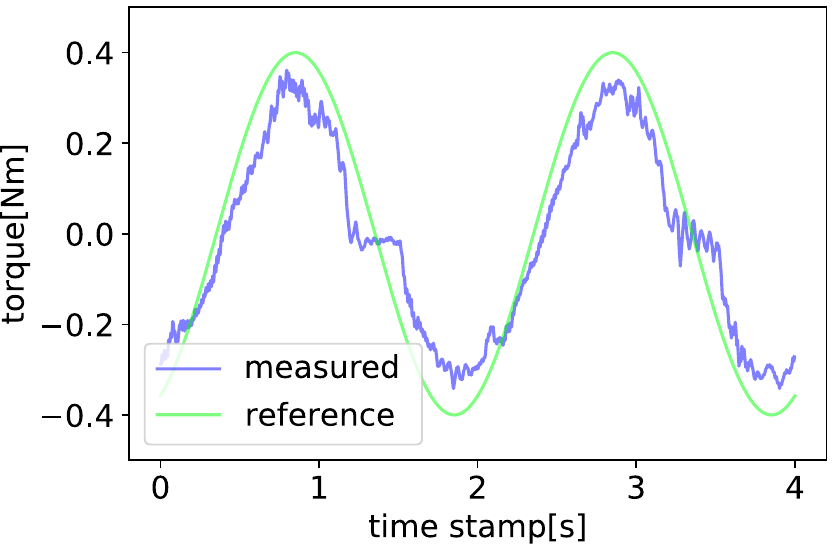}
        \subcaption{0.5Hz}
      \end{minipage} &
      \begin{minipage}[t]{0.45\hsize}
        \centering
        \includegraphics[width=\linewidth]{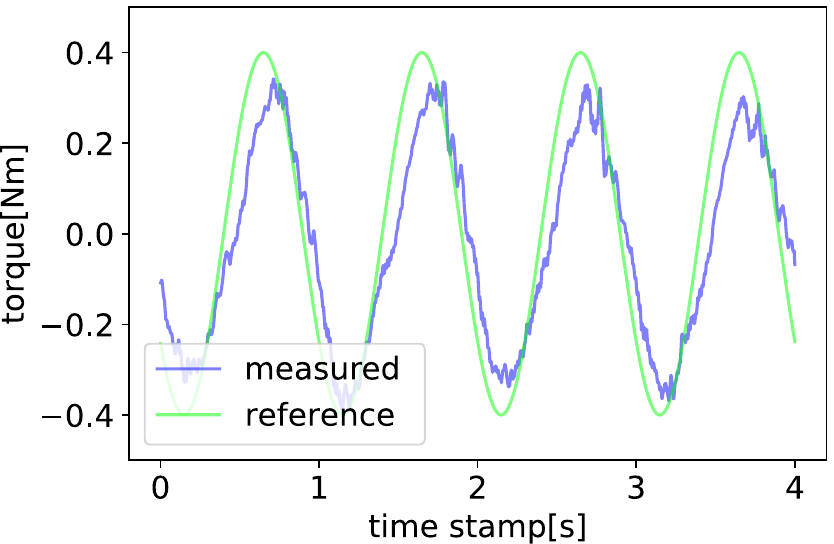}
        \subcaption{1Hz}
      \end{minipage} \\

      \begin{minipage}[t]{0.45\hsize}
        \centering
        \includegraphics[width=\linewidth]{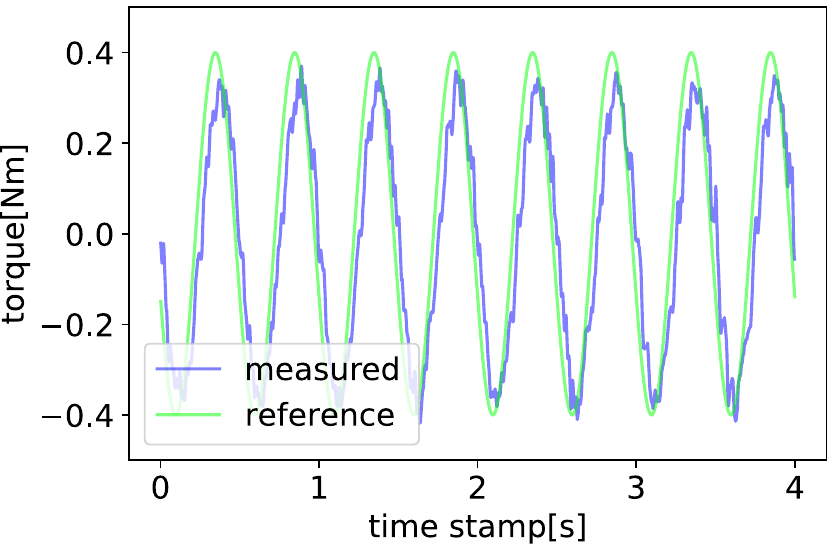}
        \subcaption{2Hz}
      \end{minipage} &
      \begin{minipage}[t]{0.45\hsize}
        \centering
        \includegraphics[width=\linewidth]{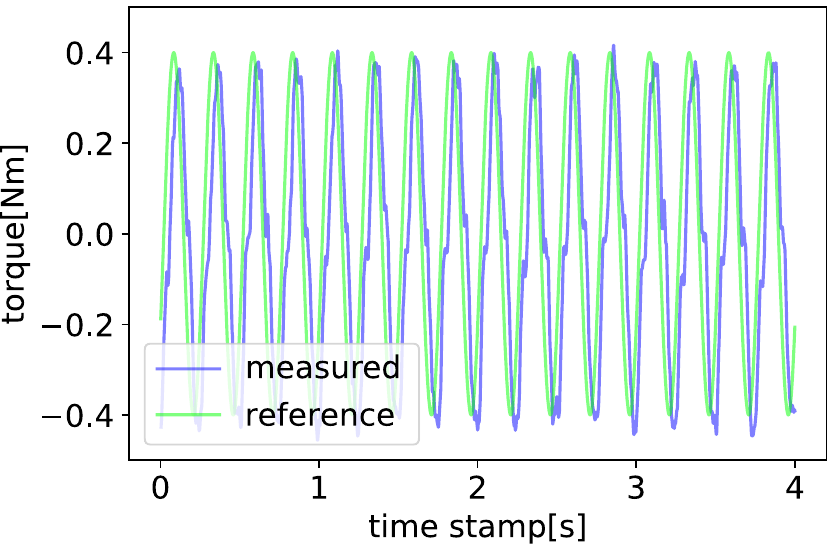}
        \subcaption{4Hz}
      \end{minipage}
    \end{tabular}
     \caption{Reference moment (green) and raw moment from the torque sensor (blue)}
    \Description{Reference moment (green) and raw moment from the torque sensor (blue)}
     \label{fig:sin_response}
  \end{figure}

A plot of the sinusoidal input moment versus the measured moment is presented in the Fig. \ref{fig:sin_response}.
As shown, the system can respond sufficiently to high-frequency target moments.
The delay for the input was approximately 60 ms, which is shorter than that of jet propellers, with an output force at a delay of approximately 300 ms \cite{heo2018thor,10.1145/3332165.3347926}.

Also, if the CMG system constantly outputs torque in the same direction, it will run out of angular momentum in that direction and will not be able to output any more torque in that direction.
This state is called saturation.
By estimating the maximum angular impulse ($torque \times time$ [\si{Nms}])  that the system can output before reaching saturation, it is possible to determine how much and how long the system can output torque.
The maximum angular impulse occurs at 0.5 Hz and its value is obtained as $\SI{0.4}{Nm}\times\SI{1.0}{s}\times 2 / \pi = \SI{0.25}{Nms}$ as the time integral of the torque for one peak of the sine function.
A value of 0.38 Nms was observed in research \cite{Winfree2010-qf}, indicating that MetamorphX can output an angular force product on the same order of magnitude.

\subsection{Evaluation in Motion : Accuracy of Impedance Control}\label{sec:evalimp}
We investigated whether the device can be integrated with an IMU to perform impedance control.
The measurement method for impedance control is described below.
The torque sensor was mounted between the handle and the torque generator, as described in the previous section.
The moment applied to the torque sensor $\tau_{sensor}$ can be expressed as follows:
\begin{equation}
    \tau_{sensor} = -I\dot\omega - D\omega +\tau_{gen} + G(\theta)
\end{equation}

The viscosity coefficient $D$ due to the contact between MetamorphX and air is considered to be zero.
The moment due to gravity $G(\theta)$ is obtained from the gravity vector $g(\theta)$ from the IMU and the position of the center of gravity $x_G$ as follows: $G(\theta) = x_G \times g(\theta)$.
Therefore, by removing the effect of the moment due to gravity from the sensor value, the moment $\tau_{imp}$ is obtained as
\begin{equation}
    \tau_{imp} = -I\dot\omega + \tau_{gen}
\end{equation}
In this situation, if moment is generated according to impedance control, then
\begin{equation}
    \tau_{gen} = -\Delta I \dot\omega-\Delta D \omega
\end{equation}
Thus, the moment is expressed as
\begin{equation}
    \tau_{imp} = -(I+\Delta I) \dot\omega -\Delta D \omega
    \label{eq:sensored_impedance}
\end{equation}

\subsubsection{Evaluation Setup}

\begin{figure}
    \centering
    \includegraphics[width=\linewidth]{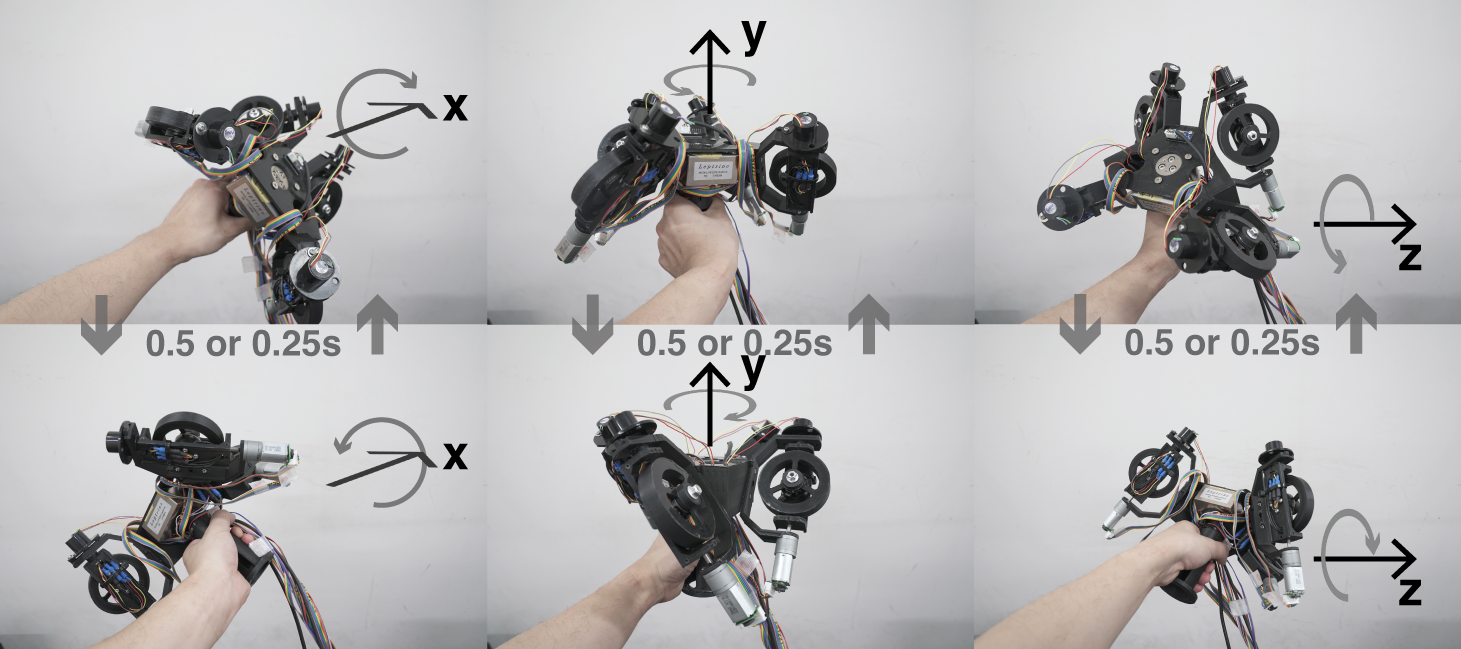}
    \caption{Directions of swinging axis. It takes 0.5 or 0.25 seconds to move to the opposite state.}
    \Description{Directions of swinging axis. It takes 0.5 or 0.25 seconds to move to the opposite state.}
    \label{fig:eval_motion}
\end{figure}

The participant was asked to hold the device and perform swinging motions in a specified direction $j$ (x-, y-, z-axis), as shown in Fig. \ref{fig:eval_motion}.

We considered human motion nearly a composite of multi-axis repetitive motion due to the limited range of motion of the joints. In other words, when a user moves their hand in a particular direction, it always returns to its original location. Hence, we conducted a study on repetitive motion along each axis.

Note that a single participant performs all motions to keep the motion trajectory as similar as possible across conditions.
To make the motions performed by the participant as identical as possible with regard to each parameter $\Delta I_j,\Delta D_j(j=x,y,z)$, the participant performed the swing motion in synchronization with a metronome sound.

Specifically, while listening to the sound of a metronome at 120 or 240 BPM, the participants were asked to perform one swing motion at one count.
Thus, each swing takes 0.5 seconds and 0.25 seconds, respectively, and the frequency of the reciprocating motion is 1 Hz and 2 Hz, respectively.

We verify that the relationship between the angular acceleration $\dot\omega$ or angular velocity $\omega$ of the motion and the moment $\tau_{imp}$ is the value given by the equation (\ref{eq:sensored_impedance}).
The initial impedance of the device was given as follows: $(I_{x},I_{y},I_{z}) = (0.0065,0.0075,0.003){\rm kgm^2}$, $(D_x,D_y,D_z) = (0,0,0){\rm Ns}$.
Note that we cannot precisely determine the viscosity coefficient because it depends on the radius of the swing motion, but it can be roughly estimated to be $O(10^{-5})$, which is sufficiently smaller than the amount we change ($O(10^{-1})$), so we treated it as 0 here.

The parameters $\Delta I_j$ and $\Delta D_j$ were varied for this initial impedance.
First, the inertia difference $\Delta I_j$ was varied.
For the motion of each axis, the change in the moment of inertia of the corresponding axis $\Delta I_j$ was set as (0, 0.005, 0.01, 0.015) \si{kgm^2}.
A total of 24 trials were conducted with 3 rotation axes and 4 parameters $\Delta I_j$ and 2 swing frequency conditions.
Next, the viscosity coefficient difference $\Delta D_j$ was varied.
For the motion of each axis, the change in the viscosity coefficient of the corresponding axis $\Delta D_j$ was set as (0, 0.05, 0.1, 0.15) \si{Ns}.
Again, 24 trials were also conducted with 3 rotation axes and 4 parameters $\Delta D_j$.
The participant performed the swing motion for 10 seconds in each condition.
There was a 1-minute break between each condition.
\subsubsection{Results: Change in Inertia}
 We analyzed whether impedance control could change the inertia.
 In Equation (\ref{eq:sensored_impedance}), $\Delta D=0$, the sensor values are given as follows:
\begin{equation}
    \tau_I = -(I+\Delta I) \dot\omega
\end{equation}

\begin{figure}
    \centering
    \includegraphics[width=\linewidth]{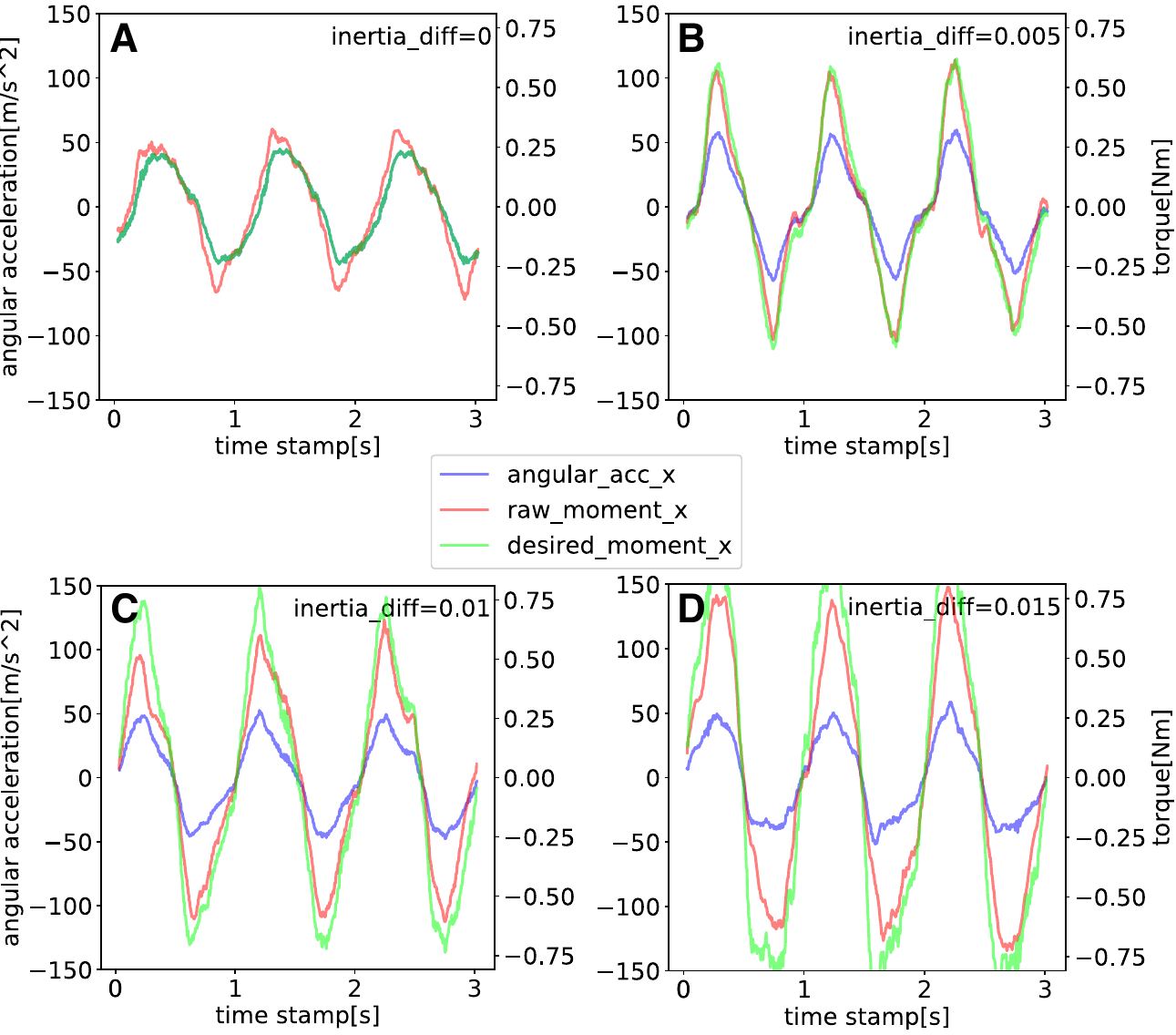}
    \caption{Angular acceleration (blue), raw moment from the torque sensor (red) and desired moment $\tau_I$ (lime) for 1Hz x-axis motion. $ \Delta I_x = \{0,0.005,0.010,0.015\}$ for \{A,B,C,D\}}
    \Description{Angular acceleration (blue), raw moment from the torque sensor (red) and desired moment $\tau_I$ (lime) for 1Hz x-axis motion. $ \Delta I_x = \{0,0.005,0.010,0.015\}$ for \{A,B,C,D\}}
    \label{fig:i-x-1Hz}
\end{figure}

\begin{figure}
    \centering
    \includegraphics[width=\linewidth]{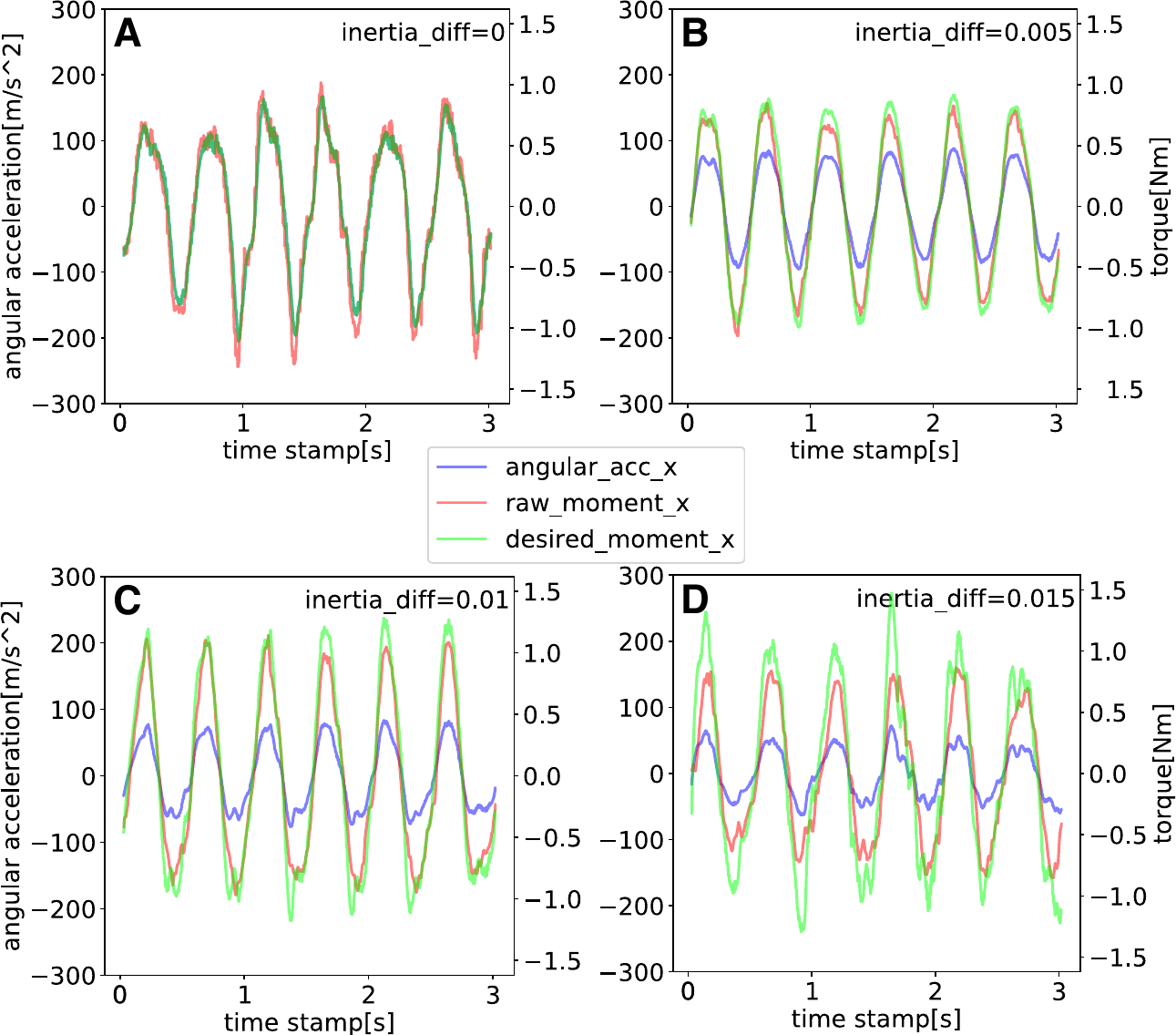}
    \caption{Angular acceleration (blue), raw moment from the torque sensor (red) and desired moment $\tau_I$ (lime) for 2Hz x-axis motion.$ \Delta I_x = \{0,0.005,0.010,0.015\}$ for \{A,B,C,D\}}
    \Description{Angular acceleration (blue), raw moment from the torque sensor (red) and desired moment $\tau_I$ (lime) for 2Hz x-axis motion.$ \Delta I_x = \{0,0.005,0.010,0.015\}$ for \{A,B,C,D\}}
    \label{fig:i-x-2Hz}
\end{figure}

From this equation, the time evolution of the sensor value is expected to be the value of angular acceleration multiplied by $-(I+ \Delta I)$.
Parts of the waveforms of the angular acceleration $\dot\omega$, raw moment, and desired moment $\tau_{I}$ of each swing frequency and impedance difference parameter are shown in Fig. \ref{fig:i-x-1Hz}, \ref{fig:i-x-2Hz}.
For readability, we only show the x-axis results in this section.
The y- and z-axis results are included in the appendix (Fig. \ref{fig:i-y-1Hz}, \ref{fig:i-y-2Hz}, \ref{fig:i-z-1Hz}, and \ref{fig:i-z-2Hz}) since they were almost identical to the x-axis results.

The scales were adjusted in the graphs to compare the similarity of the waveforms of two time-evolving values with different scales.
The scale of the axis of angular acceleration (left) multiplied by $I_j$ matched the scale of the axis of torque (right).
Then, the waveforms of the angular acceleration and torque coincided at $\Delta I_j=0$, which was consistent with the results.
\footnote{Note that when $\Delta I_x=0$ (Fig.\ref{fig:i-x-1Hz}(A) and Fig.\ref{fig:i-x-2Hz}(A)), the graphs of angular acceleration and torque coincide, resulting in a green curve with blue and lime colors mixed.}
Additionally, the larger inertia change $\Delta I_j$, the larger moment was generated for the same angular acceleration.
However, a disturbance of the moment was observed when $\Delta I_j=0.015$.
This disturbance was due to the saturation of the CMGs system when the target moment in a single swing exceeded the allowable moment.
Accordingly, the inertial difference that MetamorphX can apply for human motion is approximately $\Delta I_j \simeq $ \SI{0.01}{kgm^2}.

\subsubsection{Results: Change in Viscosity}
We examined whether the impedance control can change the viscosity coefficient.
The measured moment is proportional to the linear weighted sum of the angular acceleration and the angular velocity, according to Equation (\ref{eq:sensored_impedance}).
Because the variable is $\Delta D$ here, the moment due to the initial inertia $I$ is subtracted, and the moment $\tau_D$ in Equation (\ref{eq:D}) is evaluated.
\begin{equation}
    \tau_D = -\Delta D \omega
    \label{eq:D}
\end{equation}

\begin{figure}
    \centering
    \includegraphics[width=\linewidth]{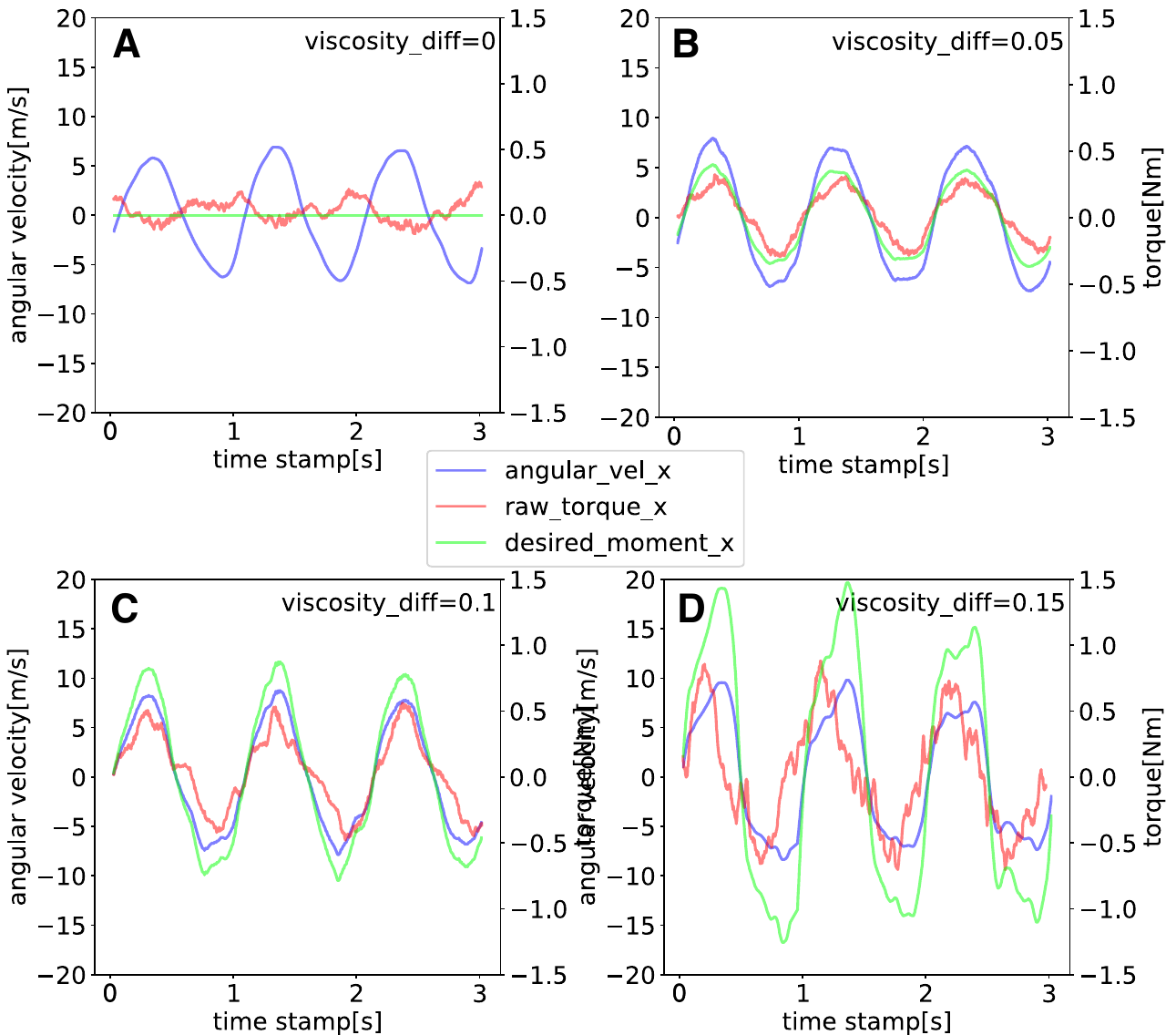}
    \caption{Angular velocity (blue), raw moment from the torque sensor (red) and desired moment $\tau_D$ (lime) for 1Hz x-axis motion. $ \Delta D_x = \{0,0.05,0.10,0.15\}$ for \{A,B,C,D\}}
    \Description{Angular velocity (blue), raw moment from the torque sensor (red) and desired moment $\tau_D$ (lime) for 1Hz x-axis motion. $ \Delta D_x = \{0,0.05,0.10,0.15\}$ for \{A,B,C,D\}}
    \label{fig:d-x-1Hz}
\end{figure}

\begin{figure}
    \centering
    \includegraphics[width=\linewidth]{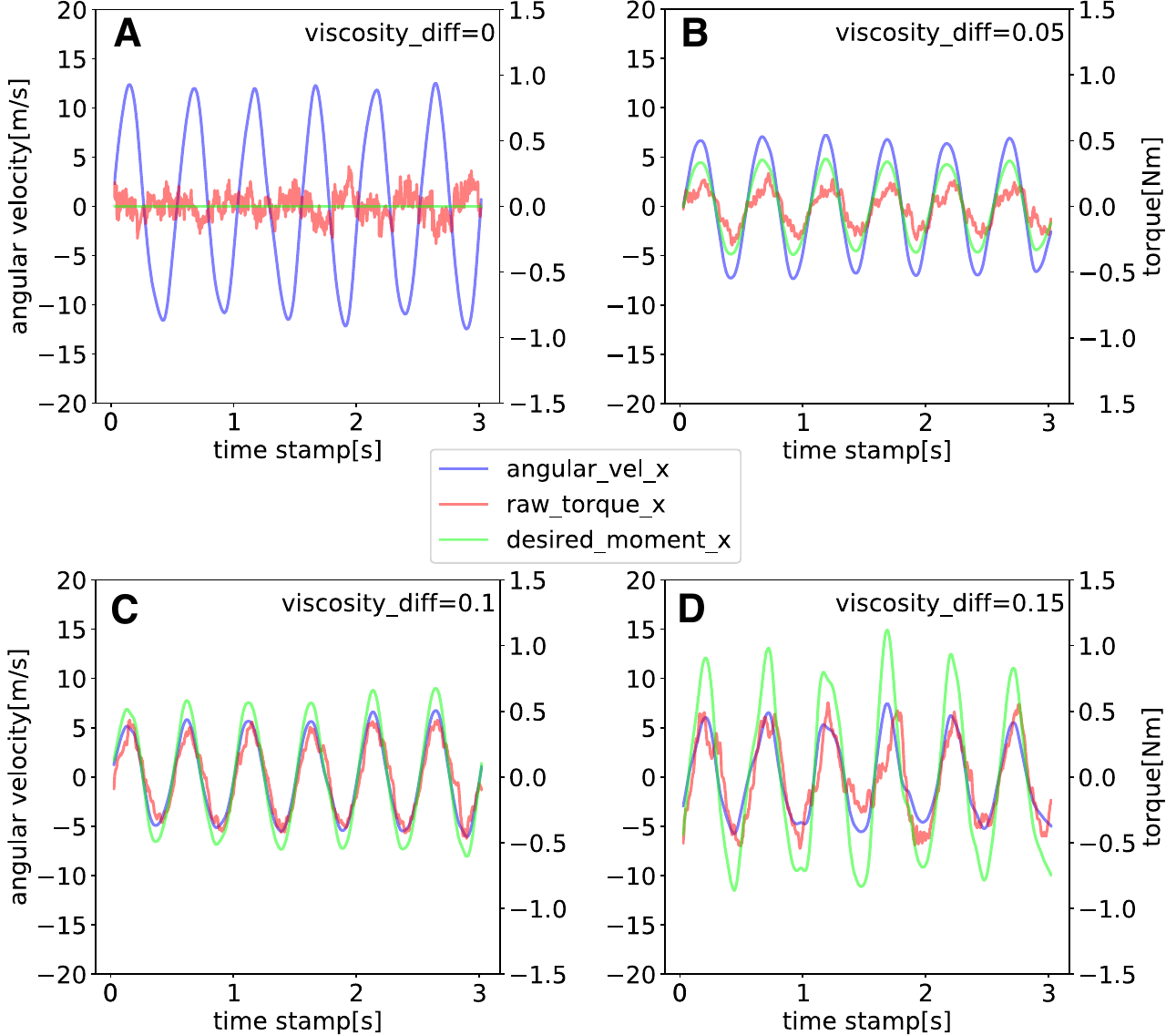}
    \caption{Angular velocity (blue), raw moment from the torque sensor (red) and desired moment $\tau_D$ (lime) for 2Hz x-axis motion. $ \Delta D_x = \{0,0.05,0.10,0.15\}$ for \{A,B,C,D\}}
    \Description{Angular velocity (blue), raw moment from the torque sensor (red) and desired moment $\tau_D$ (lime) for 2Hz x-axis motion. $ \Delta D_x = \{0,0.05,0.10,0.15\}$ for \{A,B,C,D\}}
    \label{fig:d-x-2Hz}
\end{figure}

Parts of the waveforms of the angular velocity $\omega$ and moment $\tau_{D}$ of each swing frequency and impedance difference are presented in Fig.\ref{fig:d-x-1Hz}, \ref{fig:d-x-2Hz}.
As well as the inertia results, we show only the x-axis results in this section.
The y- and z-axis results are included in the appendix (Fig. \ref{fig:d-y-1Hz}, \ref{fig:d-y-2Hz}, \ref{fig:d-z-1Hz}, and \ref{fig:d-z-2Hz}).

As shown, the assumed moment is generated in proportion to the angular velocity, in accordance with Equation (\ref{eq:D}).
Additionally, the larger viscosity change $\Delta D_j$, the larger moment was generated for the same angular velocity.
However, at $\Delta D_j=0.15$, the moment was partially not proportional to the angular velocity and disturbed.
This indicates that the system has reached moment saturation and could no longer output the moment in that direction.
According to this result, it can be said that the viscosity coefficient difference that MetamorphX can apply for human motion is approximately $\Delta D_j \simeq$ \SI{0.1}{Ns}.

After all, many readers may wonder if this device can reproduce "no forces."
In that case, we can look at the graphs of $\Delta I_j=0$ and $\Delta D_j=0$. The difference between the raw torque value (red) and desired torque value (green) is the undesired torque.
This undesired torque is small relative to the overall torque generated.
The User study session will discuss how the user feels about this undesired torque.

In this section, we showed that the inertia difference and viscosity difference that MetamorphX can present are \SI{0.01}{kgm^2} and \SI{0.1}{Ns}, respectively.
The initial impedance of the device was $(I_{x},I_{y},I_{z}) = (0.0065,0.0075,0.003)$ \si{kgm^2}, $(D_x,D_y,D_z) \simeq (0,0,0)$ \si{Ns}.
The Weber fractions of a moment of inertia and viscosity are 11-22\% \cite{Debats2012-xa} and 30-100\% \cite{Bergmann_Tiest2013-dm}, respectively, so the user can sufficiently perceive the inertia and viscosity difference represented by MetamorphX.

\subsection{Power Consumption and Noise}
In accordance with research \cite{heo2018thor,10.1145/3332165.3347926}, we evaluated the power consumption and noise generated by MetamorphX.
The power consumption is always 1.5A at 11.1V to rotate the flywheel.
In addition, the gimbal drive consumed an average of 0.5A at 12V; thus, approximately 23 W of power was needed to run the device.

The noise level was measured at a distance of 1 m from the device using a decibel meter (PT6708, Protmex) when the flywheel was rotated at 800 rad/s.
The result was 75 dB, which is lower than that of a system using a jet propeller (approximately 80-90dB).
The relatively low noise level is an advantage of using CMGs.

\section{Application}\label{sec:app}
The impedance control of MetamorphX allows the inertia and viscosity of the motion to be changed instantaneously by changing the feedback parameters $\Delta I$ and $ \Delta D$.
This allows the following representations.

\begin{itemize}
    \item Representation of multiple moments of inertia (e.g., user switches between multiple tools)
    \item Representation of multiple viscosity (e.g., reproducing drag forces that vary with respect to the shape of the tool)
    \item Representation of dynamic changes in viscosity (e.g., user moves the oars of a boat in and out of water)
    \item Representation of dynamic changes in moment of inertia (e.g., inertia decreases when a tool breaks / inertia increases when the tool adheres to an object in the environment)
\end{itemize}

We present the two following applications as examples that include all of the aforementioned representations.

\subsection{Hand-tool-interaction Playground in VR}
In this research, we focus on motion while we grasp an object.
However, when we use tools, there is a sequence of events in which an object exists in the environment, then we reach out to touch it and finally grasp it \cite{10.3389/frobt.2021.714023}.
Therefore, in this application, we consider combining Auto Hand\footnote{https://www.earnestrobot.net/unityassets/autohand}, which enables high-quality physical interaction in VR, with a system that can switch between grasping and non-grasping.
In this application, we calculate the tool's inertia tensor and viscous resistance at the hand's coordinates when grasping an object in a virtual environment.
This allows MetamorphX to reproduce the force feedback felt when grasping various objects from different directions.

\begin{figure*}
    \centering
    \includegraphics[width=\linewidth]{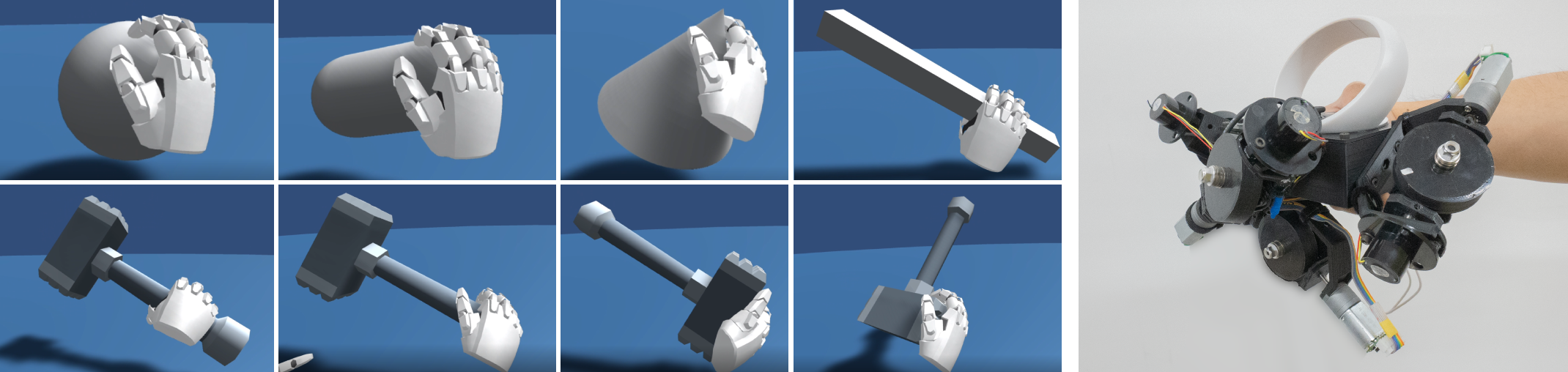}
    \caption{(Left) List of typical shapes handled by Auto Hand VR. Impedance changes with different objects and even the same object can change its impedance depending on the grasped position. (Right) A device with a built-in commercial controller for receiving user input and tracking.}
    \Description{(Left) List of typical shapes handled by Auto Hand VR. Impedance changes with different objects and even the same object can change its impedance depending on the grasped position. (Right) A device with a built-in commercial controller for receiving user input and tracking.}
    \label{fig:autohand}
\end{figure*}

Figure \ref{fig:autohand} shows a list of typical shapes handled in this application.
The impedance varies depending on the object's shape and the grasped position.
The user can grasp any part of these objects and feel the force feedback changing depending on the grasping position.
We created a device incorporating a Meta quest Touch controller to receive the user's grasping/non-grasping input.

\subsection{Campfire VR}
\begin{figure}
    \centering
    \includegraphics[width=\linewidth]{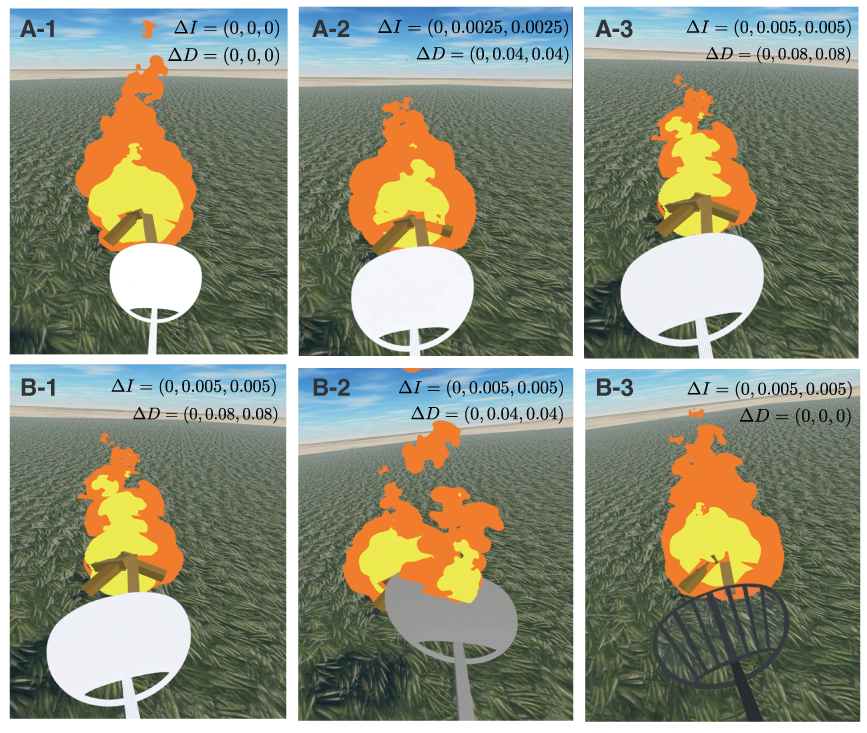}
    \caption{(A 1--3) A larger fan size corresponds to stronger wind generated by the fan is. (B 1--3) When the fan burns, the wind generated by the fan gradually becomes weaker. The change in the impedance for each state is shown in the upper left.}
    \Description{(A 1--3) A larger fan size corresponds to stronger wind generated by the fan is. (B 1--3) When the fan burns, the wind generated by the fan gradually becomes weaker. The change in the impedance for each state is shown in the upper left.}
    \label{fig:app}
\end{figure}

In this application, the user grasps a fan and interacts with a campfire by moving the fan in a virtual space.
The wind is generated when the user swings the fan perpendicular to the surface.
Conversely, no wind can be generated by swinging the fan parallel to the surface.
For stronger wind, a larger amount of drag returns to the arm.
The user can change the size of the fan (in Fig. \ref{fig:app} A 1--3).
With an increase in the fan size, stronger wind can be produced, but the fan’s inertia and viscosity increase, making it more difficult to swing.
In addition, if the fan is close to the fire, it will burn. so the fan should be kept away from the fire.
If the fan burns, its moment of inertia does not change significantly, but the viscosity coefficient decreases (in Fig. \ref{fig:app} B 1--3).
This burned fan can not generate wind.

\section{User Study}

We asked users to experience the applications described in section \ref{sec:app} to investigate how well MetamorphX reproduces inertia and viscosity.
The application to be experienced was the Campfire VR, which allows the user to experience various inertia and viscosity of grasped object.

\subsection{Method}
Nine participants (all males aged 21 to 29) experienced the Campfire VR applications described above.
Participants began the experience wearing a head-mounted display (VIVE Pro, HTC) and noise-canceling headphones (WH-1000XM4, Sony) and grasping the MetamorphX in their hands.
First, the participants experienced waving different-sized fans to send wind to the fire (Fig. \ref{fig:app} A 1--3).
Next, they experienced the fan begin to burn, and the wind and resistance created by the fan gradually decreased. (Fig. \ref{fig:app} B 1--3).
The experience ended when the wind became completely unable to wake up.
After the experience, participants answered a subjective questionnaire.
The questionnaire was a 5-point Likert scale evaluation (1: strongly disagree - 5: strongly agree), with the following five questions.
The experience concluded with a post-hoc interview.
\begin{enumerate}
    \item[Q1] Did the size of the fan change the force when you shook it?
    \item[Q2] Did you feel any change in resistance after the fan burned?
    \item[Q3] Did you feel as if a force was being exerted on you from the outside?
    \item[Q4] Did you feel that the moment was generated as a result of your movement?
    \item[Q5] Do you think the moment feedback changed the speed of your movement?
\end{enumerate}

Q1 and Q2 are questions to verify whether participants can sense the changes in the moment output through the device.
Q3 and Q4 are questions to check whether the moment is generated in response to participants' motion.
Q5 is a question to check whether participants' motion is changed by force feedback.

\subsection{Results: Questionnaire}
\begin{figure}
    \centering
    \includegraphics[width=0.8\linewidth]{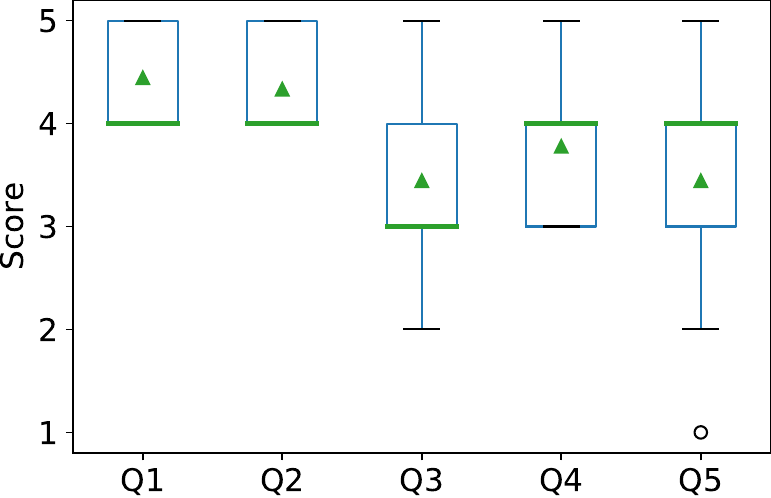}
    \caption{The results of participants' answer about our demo. The green line represents the median and the green triangle represents the mean.}
    \Description{The results of participants' answer about our demo. The green line represents the median and the green triangle represents the mean.}
    \label{fig:question}
\end{figure}

The results of questionnaire are shown in Fig. \ref{fig:question}.
High scores were obtained for both Q1 (Med: 4, IQR: 4--5) and Q2 (Med: 4, IQR: 4--5).
It suggests that humans can perceive the change in impedance represented by the device.
Although the scores for Q3 (Med: 3, IQR: 3--4) and Q4 (Med: 4, IQR: 3--4) were generally high, the results varied among the participants.
One participant noted that ``{\it When I waved the large fan, I felt uneven resistance, perhaps because the device was trying to generate the maximum output}''.
This means that the device sometimes failed to output the desired moment due to the saturation of the CMGs system.
This sometimes causes the participants to feel that the system does not respond to their own motion.
The results for Q5 (Med: 4, IQR: 3--4) were also varied among the participants.
The larger the impedance represented, the greater the force required to swing the device.
Therefore, we expected that the speed at which the device swung would change, but the results varied.
In this application, the fire flickers strongly when the wind is stronger.
It is possible that the participants dynamically moved the fan to make the fire strongly flicker.
We found that the swinging motion of a grasping device is influenced not only by haptic information but also by how the motion affects the environment.

\subsection{Results: Post-hoc Interview}
We also analyzed what the users felt during the experience based on the post-hoc interview after the experience.
We manually labeled the transcriptions of the interviews and extracted topics many users addressed.
The topics are listed below.

\subsubsection{Change in Impedance}
Six users mentioned that the size of the fan is related to the amount of inertia and
drag force they feel when they shake it.
We can see that MetamorphX presented various impedance to the users as this demo aimed.
Typical comments were, ``{\it As the fan got bigger and bigger, the wind seemed to get stronger and stronger.}'' and ``{\it I felt the fan got heavier and harder to swing as its size changed.}''

\subsubsection{Gap between Visually Expected Weight and Haptically Perceived Weight}
Four users mentioned a gap between the visually expected and haptically perceived weight.
Although the demonstration showed a paper fan, two users mentioned that they felt as if they were waving a metal fan.
Since humans visually predict the weight of an object based on its size and material, they feel strange when the force feedback differs from that prediction.
Since MetamorphX cannot change its weight, we must reduce the gap between the visual weight and the perceived weight from force feedback by adjusting the material and size of the object displayed to the user.
Typical comments were, ``{\it It was more like fanning an iron fan than a paper fan.}'', ``{\it In general, the device is much heavier than expected by looking just the visual representation.}''

\subsubsection{Unnecessary Force}\label{seq:unwanted}
Two users mentioned the sensation of an unnecessary external force while shaking the device.
They commented: ``{\it When I waved a large fan, I felt uneven resistance, like ``gurgling,'' as if I was trying to generate maximum output.}'' and ``{\it I felt a sensation of vibration when shaking a large fan.}''
Both reported feeling an unnecessary force when swinging the large fan, and this is because the maximum angular force product (\SI{0.25}{Nms}) that can be output by the device was exceeded when outputting large torque feedback in a short time.
Depending on the application, it may be necessary to adjust the value of the impedance change by estimating the maximum angular acceleration and angular velocity that can occur based on the anticipated user interaction.

\subsubsection{Difference in Torque Sensitivity depending on the Direction of Swing}
Two users mentioned that they felt the force differently depending on the direction in which they waved the object.
They commented: ``{\it It was easier to feel the force in the order of roll, yaw, and pitch.}'' and ``{\it The force sensation felt when I turned the right hand outward (the direction of rotation for the doorknob opening motion) was stronger than in the opposite direction.}''
We designed the arrangement of each CMG in MetamorphX to provide equal torque output in all three axis directions.
If there is a difference in the sensitivity of perceived force depending on the direction of motion of the human hand, We can change the placement of the CMGs accordingly to transmit torque to humans more efficiently.
We believe that will make it possible to efficiently represent objects' size, shape, and other characteristics perceived by humans over a wider range.

\subsubsection{Enjoyment}
Five users mentioned that they felt a sense of reality and enjoyment from the moment feedback and the visual representation (flickering fire) changing in sync with each other.
This suggested that the VR experience was enhanced by MetamorphX.
Typical comment was ``{\it It was a very exciting experience, including the drag force change with the size of the fan.}''

Based on the results of the above questionnaire and interviews, we found that MetamorphX was able to simulate inertia and viscosity.
However, we found some problems through the experience, such as the weight of the device and the limit of the user's motion speed.
We discuss how to deal with them in the following section.

\section{Limitations and Future Work}

\subsection{Quantitative Perceptual Experiments}

In this study, we mainly focused on technical evaluation of the device and qualitative user studies.
We conducted only a qualitative evaluation to determine how much the difference in impedance presented by MetamorphX affects human perception, and it is necessary to verify how much it changes human perception quantitatively.
Changes in the impedance of a grasped object will change the perception of its size, weight, and shape. By Following studies investigating the human perception of grasped objects \cite{shigeyama2019transcalibur}, we should clarify the perceptual representation range of MetamorphX.

We also need to investigate how the unwanted torque described in section \ref{seq:unwanted} affects the user experience.
By letting the user grasp the device without visual stimuli, we investigate the haptic sensation by interviewing the user.
We want to find out what kind of haptic representation dimensions MetamorphX has and whether it is the same as the representation dimensions we assume (size, weight, shape).
In addition, we conducted the user study during a pandemic situation, so we could only gather participants from limited backgrounds on campus.
To observe effects independent of age and gender, we will conduct future experiments with participants from a wider range of backgrounds.

\subsection{Weight of the Device}

Although the device is equipped with a mechanism to generate a moment, it is much heavier than the controllers used in typical VR experiences.
In the feedback on the application experience, some participants stated that the device was so heavy that it was sometimes difficult to feel the generated moment.
If the device can be made lighter, the moment output from the device will be more easily felt by the users.

MetamorphX weighs 840g, and its mass composition is as follows: flywheel motor 120 g (14\%), gimbal motor 170 g (21\%), flywheel 120 g (14\%), mechanical parts (bearings and shaft) 120 g (14\%), and frame 310 g (36\%).
The frame is made of PLA, and its thickness is 5 mm for rigidity, which makes the device heavy.
If a metallic material with a higher strength/weight ratio than PLA is used, the device would be lighter while keeping the same rigidity.

\subsection{Operating Noise}
The noise level generated by the device in the technical evaluation session was measured as approximately 75 dB, which is comparable to that of a vacuum cleaner.
In practice, noise-canceling earphones are considered to reduce the effect of noise on humans as much as possible.
During the user study, users wore noise-canceling headphones, and while some mentioned the device's vibration during the post-hoc interview, no one mentioned the noise.

In addition, the flywheel was left bare to reduce the weight of the device; however, it would be expected that covering flywheels will also reduce noise.
This is an advantage of using CMGs since similar noise problems occur in systems that use jet-propeller, but covering the propellers makes it impossible to output force.

\subsection{Moment Saturation}
As observed in the technical evaluation and application session, the CMGs system may cause moment saturation when attempting to output large moments in a short period.
This saturation is unavoidable when CMG techniques are used.
Therefore, it would be necessary to estimate in advance the maximum velocity and the maximum acceleration of the target motion and adjust the value of the change in the impedance accordingly, so that the CMG does not fall into a saturated state.

In addition, it is difficult to continue generating a moment that is always applied in the same direction, such as the moment of gravity.
One possible solution is to use a CMG with variable flywheel speeds (VSCMG) \cite{schaub1998feedback}.
This allows the angular momentum of the gimbal to vary, avoiding saturation conditions.
However, using a VSCMG makes the control more complex; thus, there is a tradeoff.

\subsection{Design Policy of the Hardware}
We suppose it is possible to design a device according to the desired moment scale.
The size of the flywheel and angular velocity determines the scale of generated moment.
By estimating the values of impedance, angular acceleration, and angular velocity of desired motion in advance, the moment required to represent that motion can be calculated.
By determining the variables of the flywheel based on this calculated moment, the device can output just enough moment for the application.
In the future, we intend to develop a design theory for devices that can respond to motions of other body parts (e.g., fingers and legs).

\section{Conclusion}
This paper presents MetamorphX -- an ungrounded 3-DoF moment display system that represents various inertia and the viscosity of the motion.
Hardware employing CMGs was fabricated and connected to an IMU to generate feedback linked to human motion.
A technical evaluation and user study indicated that the proposed device can represent various inertia and viscosity, enhancing the VR experience.

\begin{acks}
This work was supported by JSPS KAKENHI Grant Number JP21J12372, 20K21801, 22H03628.
We thank Reo Matsumura in Karakuri products\footnote{https://krkrpro.com/} for his grateful advice on the device design.
We thank all participants and reviewers for their feedback, especially during the COVID-19 pandemic.
\end{acks}

\bibliographystyle{ACM-Reference-Format}
\bibliography{ref}
\appendix
\renewcommand\thefigure{\thesection.\arabic{figure}}
\section{Additional Result of Technical Evaluation}
\setcounter{figure}{0}

\begin{figure}[h]
    \centering
    \includegraphics[width=\linewidth]{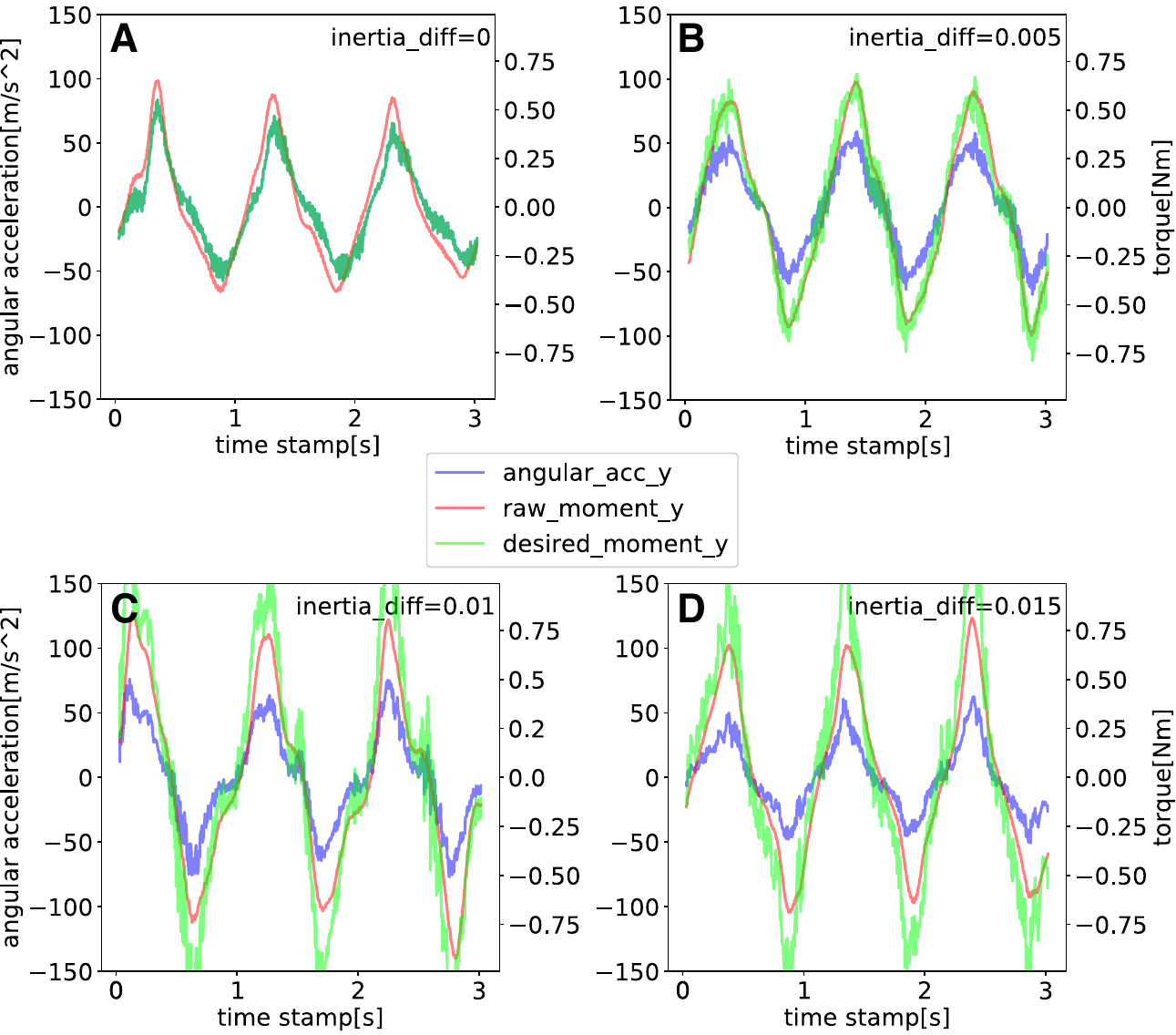}
    \caption{Angular acceleration (blue), raw moment from the torque sensor (red) and desired moment $\tau_I$ (lime) for 1Hz y-axis motion. $ \Delta I_y = \{0,0.005,0.010,0.015\}$ for \{A,B,C,D\}}
    \Description{Angular acceleration (blue), raw moment from the torque sensor (red) and desired moment $\tau_I$ (lime) for 1Hz y-axis motion. $ \Delta I_y = \{0,0.005,0.010,0.015\}$ for \{A,B,C,D\}}
    \label{fig:i-y-1Hz}
\end{figure}

\begin{figure}[h]
    \centering
    \includegraphics[width=\linewidth]{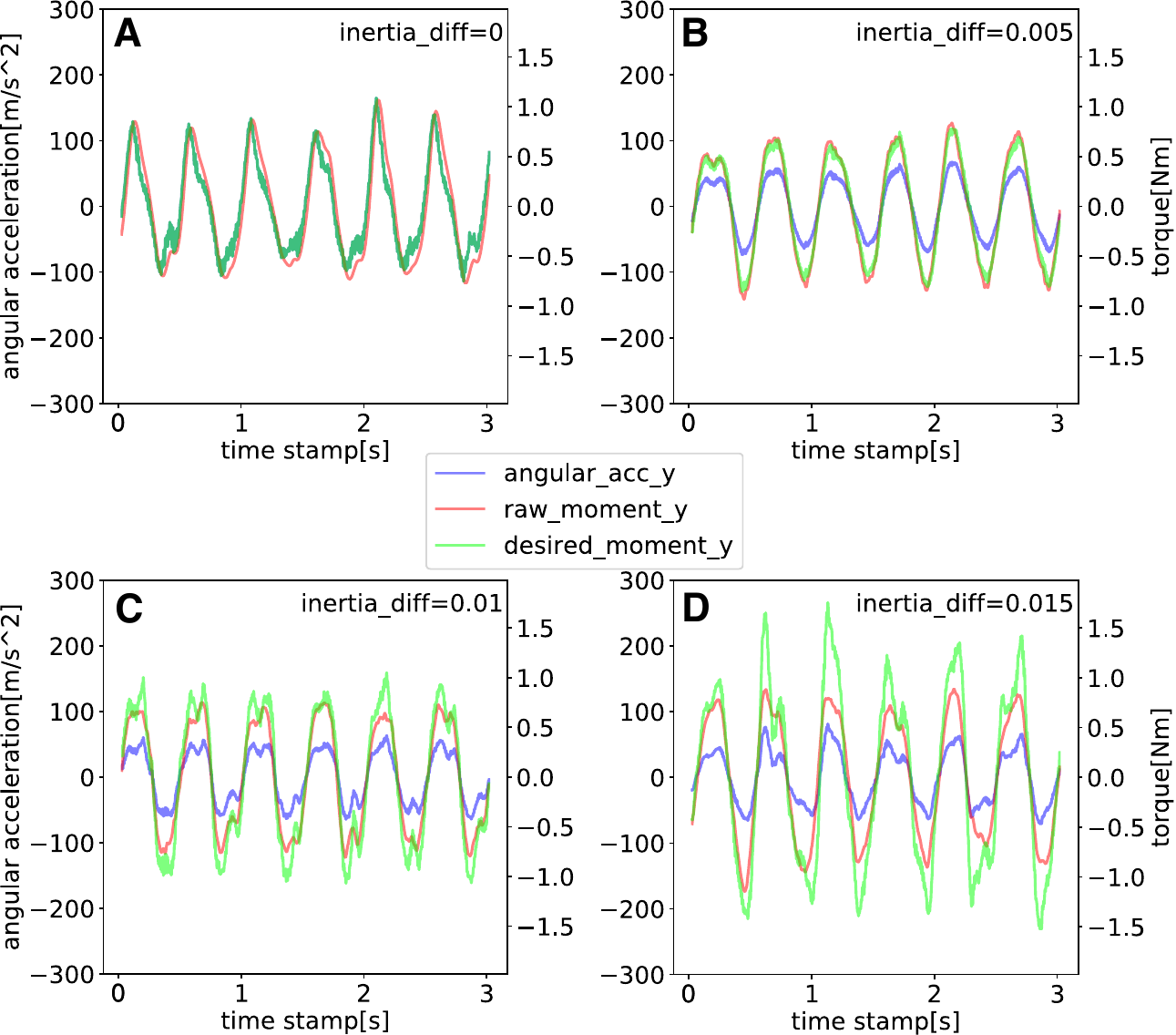}
    \caption{Angular acceleration (blue), raw moment from the torque sensor (red) and desired moment $\tau_I$ (lime) for 2Hz y-axis motion. $ \Delta I_y = \{0,0.005,0.010,0.015\}$ for \{A,B,C,D\}}
    \Description{Angular acceleration (blue), raw moment from the torque sensor (red) and desired moment $\tau_I$ (lime) for 2Hz y-axis motion. $ \Delta I_y = \{0,0.005,0.010,0.015\}$ for \{A,B,C,D\}}
    \label{fig:i-y-2Hz}
\end{figure}

\begin{figure}[h]
    \centering
    \includegraphics[width=\linewidth]{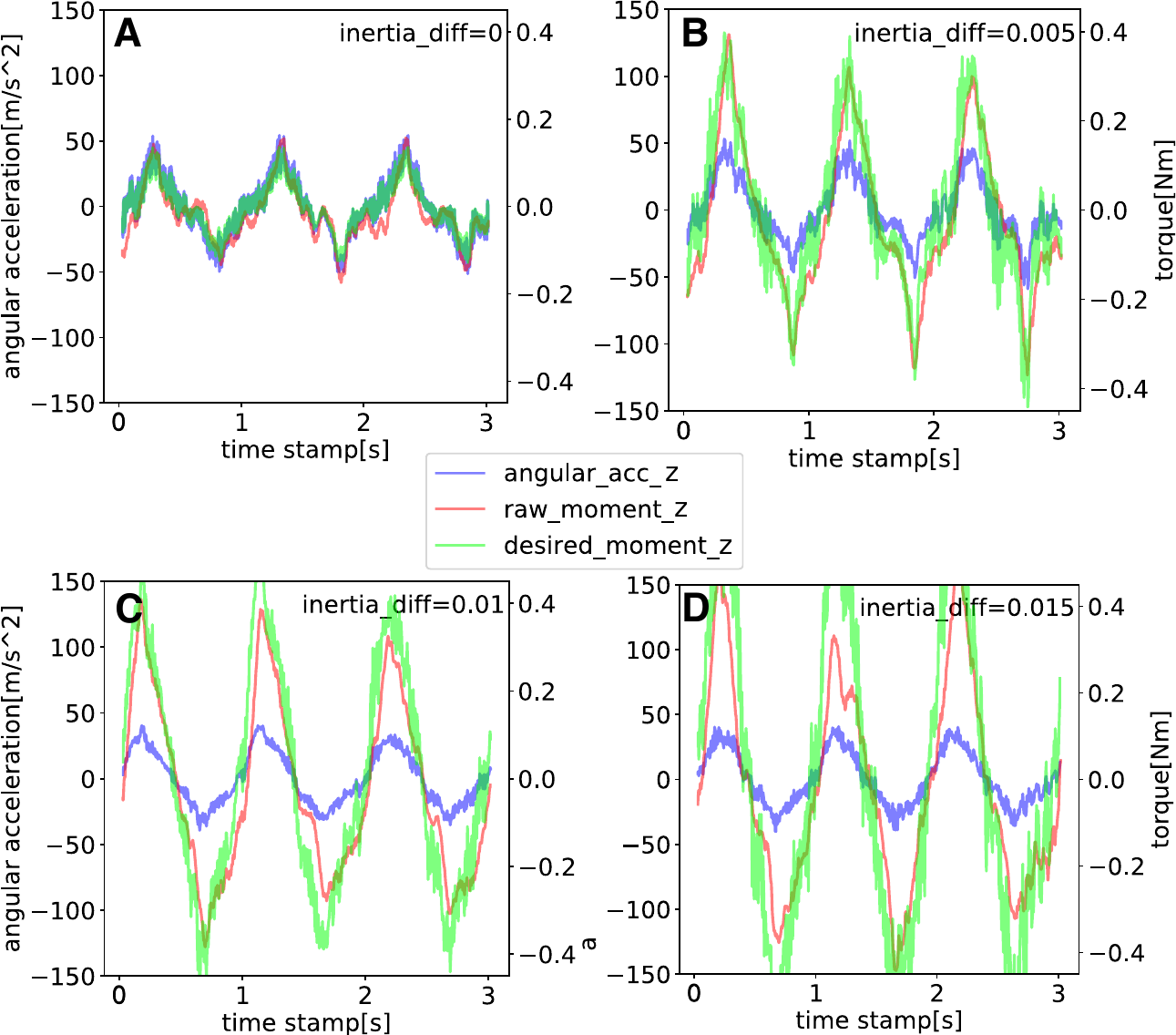}
    \caption{Angular acceleration (blue), raw moment from the torque sensor (red) and desired moment $\tau_I$ (lime) for 1Hz z-axis motion. $ \Delta I_z = \{0,0.005,0.010,0.015\}$ for \{A,B,C,D\}}
    \Description{Angular acceleration (blue), raw moment from the torque sensor (red) and desired moment $\tau_I$ (lime) for 1Hz z-axis motion. $ \Delta I_z = \{0,0.005,0.010,0.015\}$ for \{A,B,C,D\}}
    \label{fig:i-z-1Hz}
\end{figure}

\begin{figure}[h]
    \centering
    \includegraphics[width=\linewidth]{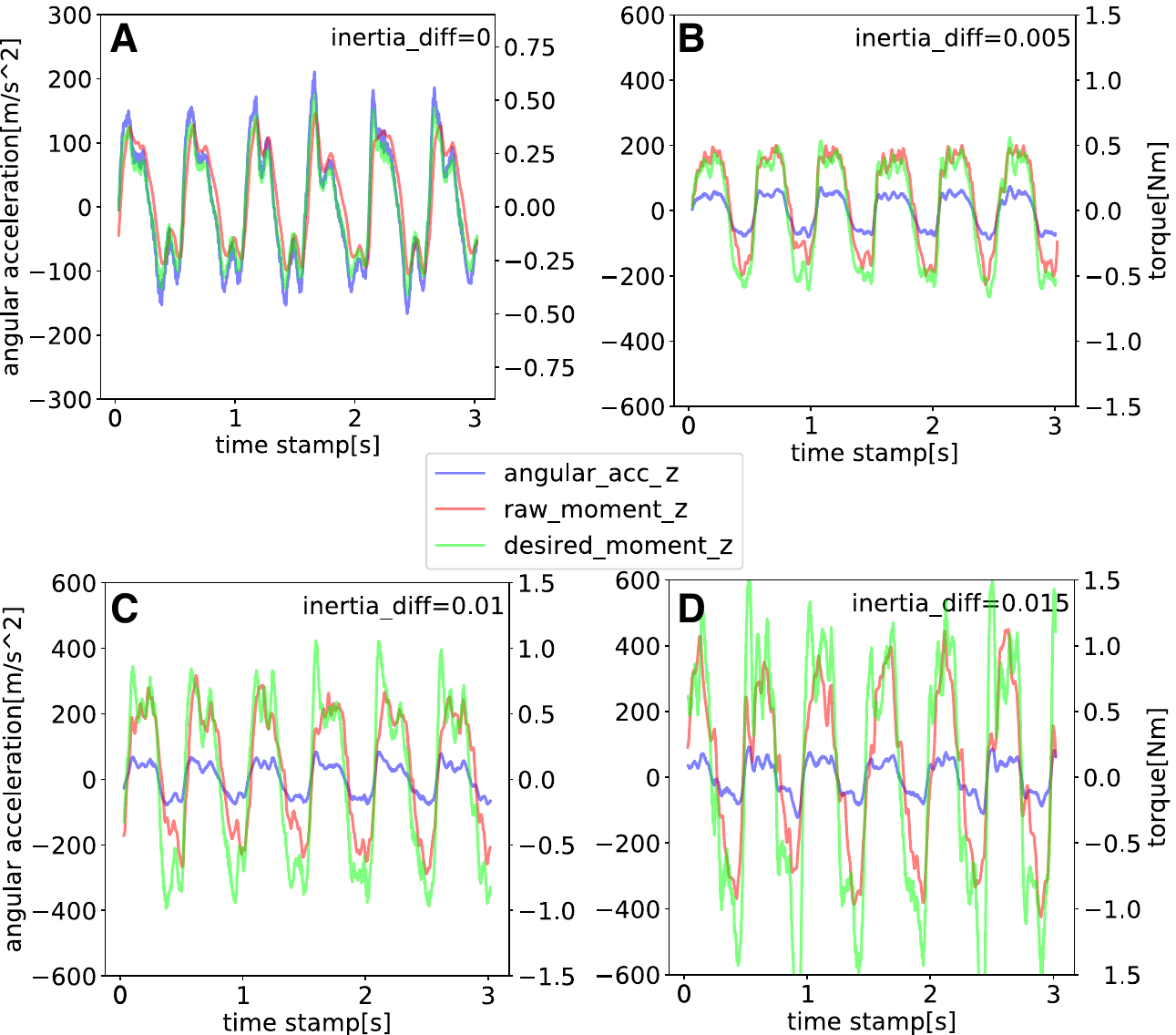}
    \caption{Angular acceleration (blue), raw moment from the torque sensor (red) and desired moment $\tau_I$ (lime) for 2Hz z-axis motion. $ \Delta I_z = \{0,0.005,0.010,0.015\}$ for \{A,B,C,D\}}
    \Description{Angular acceleration (blue), raw moment from the torque sensor (red) and desired moment $\tau_I$ (lime) for 2Hz z-axis motion. $ \Delta I_z = \{0,0.005,0.010,0.015\}$ for \{A,B,C,D\}}
    \label{fig:i-z-2Hz}
\end{figure}

\begin{figure}[h]
    \centering
    \includegraphics[width=\linewidth]{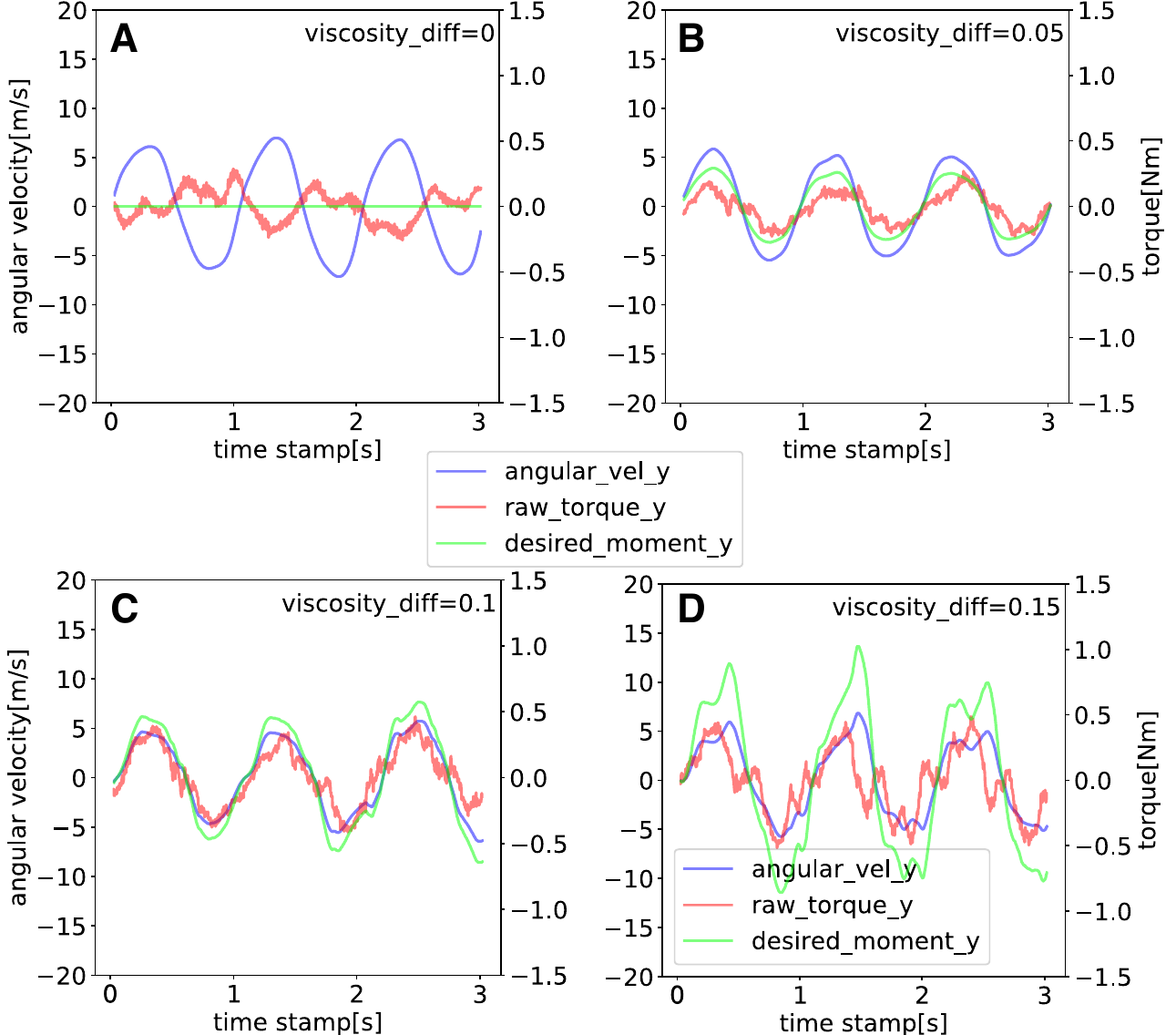}
    \caption{Angular velocity (blue), raw moment from the torque sensor (red) and desired moment $\tau_D$ (lime) for 1Hz y-axis motion.$ \Delta D_y = \{0,0.05,0.10,0.15\}$ for \{A,B,C,D\}}
    \Description{Angular velocity (blue), raw moment from the torque sensor (red) and desired moment $\tau_D$ (lime) for 1Hz y-axis motion.$ \Delta D_y = \{0,0.05,0.10,0.15\}$ for \{A,B,C,D\}}
    \label{fig:d-y-1Hz}
\end{figure}

\begin{figure}[h]
    \centering
    \includegraphics[width=\linewidth]{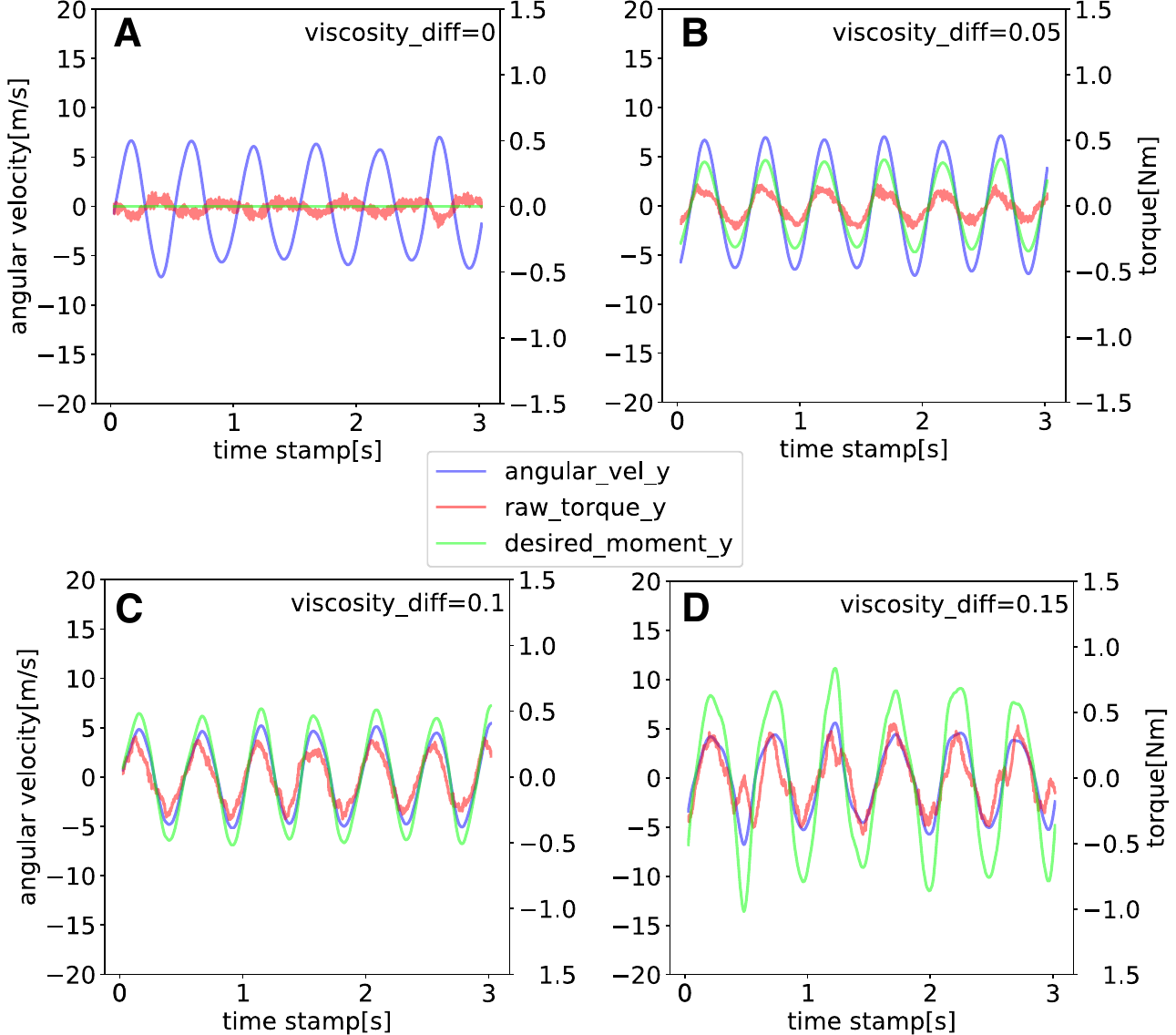}
    \caption{Angular velocity (blue), raw moment from the torque sensor (red) and desired moment $\tau_D$ (lime) for 2Hz y-axis motion.$ \Delta D_y = \{0,0.05,0.10,0.15\}$ for \{A,B,C,D\}}
    \Description{Angular velocity (blue), raw moment from the torque sensor (red) and desired moment $\tau_D$ (lime) for 2Hz y-axis motion.$ \Delta D_y = \{0,0.05,0.10,0.15\}$ for \{A,B,C,D\}}
    \label{fig:d-y-2Hz}
\end{figure}

\begin{figure}[h]
    \centering
    \includegraphics[width=\linewidth]{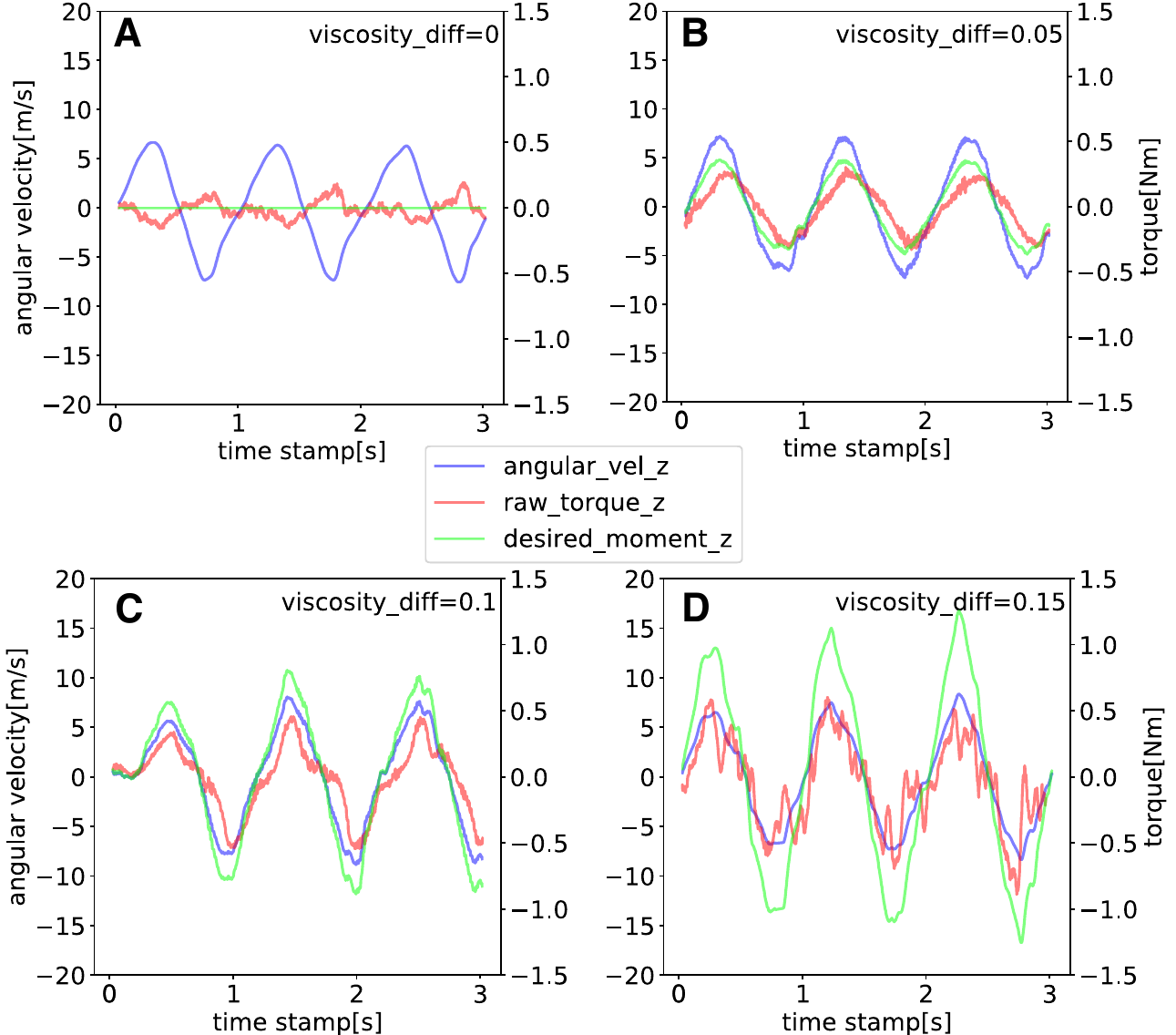}
    \caption{Angular velocity (blue), raw moment from the torque sensor (red) and desired moment $\tau_D$ (lime) for 1Hz z-axis motion.$ \Delta D_z = \{0,0.05,0.10,0.15\}$ for \{A,B,C,D\}}
    \Description{Angular velocity (blue), raw moment from the torque sensor (red) and desired moment $\tau_D$ (lime) for 1Hz z-axis motion.$ \Delta D_z = \{0,0.05,0.10,0.15\}$ for \{A,B,C,D\}}
    \label{fig:d-z-1Hz}
\end{figure}

\begin{figure}[h]
    \centering
    \includegraphics[width=\linewidth]{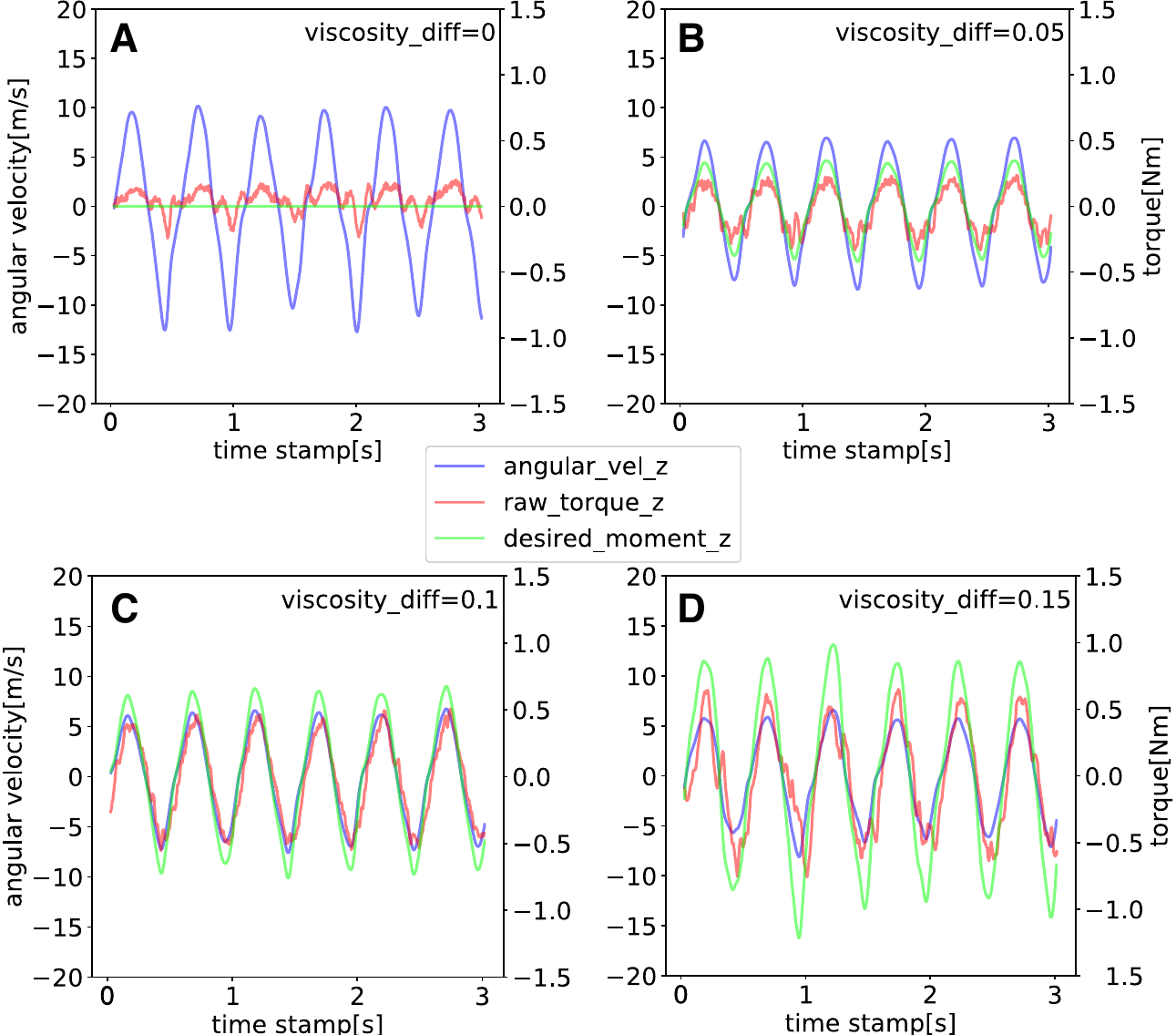}
    \caption{Angular velocity (blue), raw moment from the torque sensor (red) and desired moment $\tau_D$ (lime) for 2Hz z-axis motion.$ \Delta D_z = \{0,0.05,0.10,0.15\}$ for \{A,B,C,D\}}
    \Description{Angular velocity (blue), raw moment from the torque sensor (red) and desired moment $\tau_D$ (lime) for 2Hz z-axis motion.$ \Delta D_z = \{0,0.05,0.10,0.15\}$ for \{A,B,C,D\}}
    \label{fig:d-z-2Hz}
\end{figure}

The graphs during motion on the y and z axes, described in the technical evaluation section, are shown in this section.
The results for each condition of motion frequency (1,2 Hz), motion axis (y, z-axis), and inertial difference parameter ($\Delta I_j$ = \{0, 0.005, 0.01, 0.015\} \si{kgm^2}) are shown in Fig. \ref{fig:i-y-1Hz}, \ref{fig:i-y-2Hz}, \ref{fig:i-z-1Hz}, \ref{fig:i-z-2Hz}.
And, the results for each condition of motion frequency (1,2 Hz), motion axis (y, z-axis), and viscosity difference parameter ($\Delta D_j$ = \{0, 0.05, 0.1, 0.15\} \si{Ns}) are shown in Fig. \ref{fig:d-y-1Hz}, \ref{fig:d-y-2Hz}, \ref{fig:d-z-1Hz}, \ref{fig:d-z-2Hz}.

\newpage

\clearpage

\newpage

\end{document}